\documentclass[conference]{IEEEtran}%
\pdfoutput=1

\usepackage[papersize={8.5in,11in},verbose,footskip=0.6cm,bmargin=19.1mm,rmargin=19.1mm,lmargin=19.1mm,headsep=0.4cm,tmargin=19.1mm]{geometry}

\usepackage{amsfonts}
\usepackage{amsmath}
\usepackage{amssymb}
\usepackage{graphicx}
\usepackage{amsthm}%
\providecommand{\U}[1]{\protect\rule{.1in}{.1in}}
\newtheoremstyle{example}{\topsep}{\topsep}
{}
{}
{\bfseries}
{}
{  }
{\thmname{#1}\thmnumber{ #2}. \thmnote{ (#3)}}
\newtheorem{theorem}{Theorem}

\newtheorem{conjecture}[theorem]{Conjecture}

\newtheorem{definition}{Definition}

\newtheorem{lemma}{Lemma}

\newtheorem{proposition}[theorem]{Proposition}

\theoremstyle{example}

\newcommand{\Tr}{{\rm Tr}\;}

\newcommand{\nc}{\newcommand}
\nc{\rnc}{\renewcommand}
\nc{\beq}{\begin{equation}}
\nc{\eeq}{{\end{equation}}}
\nc{\beqa}{\begin{eqnarray}}
\nc{\eeqa}{\end{eqnarray}}
\nc{\lbar}[1]{\overline{#1}}
\nc{\bra}[1]{\langle#1|}
\nc{\ket}[1]{|#1\rangle}
\nc{\ketbra}[2]{|#1\rangle\!\langle#2|}
\nc{\braket}[2]{\langle#1|#2\rangle}
\nc{\proj}[1]{| #1\rangle\!\langle #1 |}
\nc{\avg}[1]{\langle#1\rangle}
\nc{\smfrac}[2]{\mbox{$\frac{#1}{#2}$}}
\nc{\tr}{\operatorname{tr}}
\nc{\tracedist}[1]{\Delta_{}\!\left( #1 \right)}
\nc{\fid}[1]{F\!\left( #1 \right)}

\nc{\ox}{\otimes}
\nc{\dg}{\dagger}
\nc{\dn}{\downarrow}
\nc{\cA}{{\cal A}}
\nc{\cB}{{\cal B}}
\nc{\cC}{{\cal C}}
\nc{\cD}{{\cal D}}
\nc{\cE}{{\mathcal E}}
\nc{\cF}{{\cal F}}
\nc{\cG}{{\cal G}}
\nc{\cH}{{\cal H}}
\nc{\cI}{{\cal I}}
\nc{\cJ}{{\cal J}}
\nc{\cK}{{\cal K}}
\nc{\cL}{{\cal L}}
\nc{\cM}{{\cal M}}
\nc{\cN}{{\cal N}}
\nc{\cO}{{\cal O}}
\nc{\cP}{{\cal P}}
\nc{\cR}{{\cal R}}
\nc{\cS}{{\cal S}}
\nc{\cT}{{\cal T}}
\nc{\cU}{{\cal U}}
\nc{\cX}{{\cal X}}
\nc{\cZ}{{\cal Z}}

\nc{\entI}{{\bf I}}
\nc{\entIarrow}{{\bf I}^{\leftarrow}}
\nc{\entH}{{\bf H}}
\nc{\entS}{{\bf S}}
\nc{\entHmin}{H_{\min}}

\nc{\entF}{{\bf E}_f}

\nc{\isom}{\simeq}

\nc{\rank}{\operatorname{rank}}
\nc{\rar}{\rightarrow}
\nc{\lrar}{\longrightarrow}
\nc{\polylog}{\operatorname{polylog}}
\nc{\poly}{\operatorname{poly}}
\nc{\1}{{\openone}}

\nc{\weight}{\textbf{w}}
\nc{\hamdist}{d_{H}}

\def\U{\Upsilon}

\nc{\Sp}{{{\mathbb S}}}
\nc{\RR}{{{\mathbb R}}}
\nc{\CC}{{{\mathbb C}}}
\nc{\FF}{{{\mathbb F}}}
\nc{\NN}{{{\mathbb N}}}
\nc{\ZZ}{{{\mathbb Z}}}
\nc{\PP}{{{\mathbb P}}}
\nc{\QQ}{{{\mathbb Q}}}
\nc{\UU}{{{\mathbb U}}}
\nc{\OO}{{{\mathbb O}}}
\nc{\EE}{{{\mathbb E}}}
\nc{\id}{{\operatorname{id}}}

\nc{\qubitchannel}{\id_2}
\nc{\bitchannel}{\overline{\id}_2}

\nc{\be}{\begin{equation}}
\nc{\ee}{\end{equation}}
\nc{\bea}{\begin{eqnarray}}
\nc{\eea}{\end{eqnarray}}
\nc{\<}{\langle}
\rnc{\>}{\rangle}
\nc{\Hom}[2]{\mbox{Hom}(\CC^{#1},\CC^{#2})}
\nc{\rU}{\mbox{U}}

\nc{\ob}[1]{#1}

\def\mcal{\mathcal}

\def\HKR{\mathcal{R}_{\text{HK}}}

\def\bbR{\mathbb{R}}

\def\MACone{MAC$_1$ }
\def\MACtwo{MAC$_2$ }

\def\PIavg{\Pi^n_{\bar{\rho},\delta}}
\def\PIone{\Pi^n_{m_1}}
\def\PIonepr{\Pi^n_{m_1^{\prime}}}
\def\PItwo{\Pi^n_{m_2}}

\def\PIonetwo{\Pi^n_{m_1,m_2}}
\def\PIoneprtwo{\Pi^n_{m_1^{\prime},m_2}}
\def\PIonetwopr{\Pi^n_{m_1,m_2^{\prime}}}

\usepackage{ifthen}
\newboolean{WITHHMINS}
\setboolean{WITHHMINS}{false}

\newboolean{WITHAPDX}
\setboolean{WITHAPDX}{true}
\def\ifthen#1#2{\ifthenelse{#1}{#2}{} }

\begin{document}

	\title{\vspace*{6.3mm}Quantum interference channels} 

\author{
	\IEEEauthorblockN{
	Ivan Savov\IEEEauthorrefmark{1},
	Omar Fawzi\IEEEauthorrefmark{1}, 
	Mark M. Wilde\IEEEauthorrefmark{1}, 
	Pranab Sen\IEEEauthorrefmark{1}\IEEEauthorrefmark{2},
	and 
	Patrick Hayden\IEEEauthorrefmark{1}
	 \\ \ \\
	} 
	\IEEEauthorblockA{\IEEEauthorrefmark{1}
		School of Computer Science, McGill University,  \textit{Montr\'eal, Qu\'ebec, Canada} \\
	} 
	\IEEEauthorblockA{\IEEEauthorrefmark{2}
	School of Technology and Computer Science, Tata Institute of Fundamental Research, \textit{Mumbai, India}
	}
	\vspace{-7mm}
}


	\maketitle
	\begin{abstract}
	
		The discrete memoryless interference channel is modelled as a  
		conditional probability distribution with two outputs depending on two inputs
		and has widespread applications in practical communication scenarios.
		In this paper, we introduce and study the quantum interference channel, a 
		generalization of a two-input, two-output memoryless
		channel to the setting of quantum Shannon theory. 
		%
		 %
		%
		We discuss three different coding strategies
		and obtain corresponding achievable rate regions 
		for quantum interference channels.
		%
		%
		We calculate the capacity regions in the special cases of ``very strong'' and ``strong'' interference.
		The achievability proof in the case of  ``strong'' interference
		exploits a novel quantum simultaneous decoder for two-sender
		quantum multiple access channels.
		We formulate a conjecture regarding the existence of a quantum simultaneous decoder
		in the three-sender case and use it to state the rates 
		achievable by a quantum Han-Kobayashi strategy.
		%
		%
		

		
		
	\end{abstract}

\section{Introduction}

	%
		
	
	Modern communication systems usually approach the problem of inter-carrier interference 
	by treating the interfering signals as noise.
	Indeed, techniques like code division multiple access aim to make the encoded signals 
	as similar to background noise as possible by spreading the signal power over
	large sections of the spectrum.
	%
	%
	%
	Rather than treating the interference as noise, a receiver could
	instead try to decode the interfering signals and then ``subtract'' them
	from the received signal in order to reduce (or even remove) the interference.
	%
	The development of these ideas into practical codes for $M$-user interference channels 
	would have profound implications for many areas of communications engineering.

	The theory of this problem has been studied for more than 30 years,
	in particular for channels with two senders and two receivers \cite{Sato77,Carleial78}.
	The approach of completely decoding the interfering messages applies to 
	channels with ``very strong'' interference, and it is optimal for this class of channels \cite{carleial1975case}.
	%
	For an arbitrary interference channel, it may only be possible to \emph{partially}
	decode the interfering signal. 
	Still, the receivers can achieve better communication rates using this side information
	when decoding the messages intended for them.
	The best achievable rate region for the general interference
	channel is based on partial decoding
	of the interference and is due to Han and Kobayashi \cite{HK81}.

	In this paper, we apply and extend some insights from classical information theory
	to the study of quantum interference channels (QICs).
	These channels can model physical systems such as fibre-optic cables and 
	free space optical communication channels, when operating in low-power regimes \cite{GSW11bosonic}.
	Inspired by results like
	the Holevo-Schumacher-Westmoreland theorem on the classical capacity of point-to-point channels 
	\cite{ieee1998holevo,PhysRevA.56.131}, and Winter's results on the capacity of
	quantum multiple access channels \cite{winter2001capacity}, 
	we propose the study of classical communication over quantum interference channels.

	%

		


	We structure this paper as follows. 
	In Section~\ref{sec:summary-of-results} we review our main results.
	Section~\ref{sec:prelim} introduces notation and defines the key concepts.
	In Section~\ref{sec:MAC} we discuss the quantum multiple access channel,
	and the difference between successive decoding, simultaneous decoding 
	and rate-splitting approaches to achieving the capacity.
	%
	%
	Section~\ref{sec:QIC} presents our results on the quantum interference channel.
	We conclude by stating open problems in Section \ref{sec:discussion}.
	

\section{Summary of results}
\label{sec:summary-of-results}

	We initiate the study of quantum
	interference channels, a fundamental problem of multiuser 
	communication theory. 
	As first steps in this study, we prove the capacity region
	for channels with ``very strong''  interference (Theorem~\ref{thm:carleial})
	and channels with ``strong'' interference (Theorem~\ref{thm:strong-int}).
	For general interference channels we obtain a quantum analogue of
	Sato's outer bound (Theorem \ref{thm:sato-weaker}) 
	%
	%
	and an achievable rate region inspired by the Han-Kobayashi coding strategy \cite{HK81}
	and rate-splitting \cite{GRUW01}.
	Our work serves to highlight the importance of 	quantum simultaneous decoding
	for the multiple access channel as a key ingredient for the construction of the interference channel codes.
	%
	Prior results on quantum multiple access channels are 
	based on successive decoding and time-sharing \cite{winter2001capacity}, 
	but in Theorem~\ref{thm:sim-dec-two-sender} we show that 
	a quantum simultaneous decoder exists for multiple access channels 
	with two senders.
	The quantum Han-Kobayashi coding strategy (Theorem~\ref{thm:quantum-HK-region}) requires 
	the use of a quantum simultaneous decoder for multiple access channels  with three senders.
	It is not obvious how to extend the techniques used to prove 
	Theorem~\ref{thm:sim-dec-two-sender} to the three-sender case.
	We formulate Conjecture \ref{conj:sim-dec} concerning the existence of a quantum 
	simultaneous decoder for three-sender quantum multiple access channels.
	A proof of this conjecture would have profound consequences for multiuser
	quantum information theory since it would allow for many classical information theory
	results based on simultaneous 
	decoding to be adapted to the quantum setting.
	%
	
	%
	
	%

\section{Preliminaries}
\label{sec:prelim}

	In this section, we define the quantum interference channel
	and the communication task that we are trying to achieve.

	\subsubsection{Notation}

		We denote quantum systems as $A$, $B$, and $C$ and their corresponding Hilbert
		spaces as $\mathcal{H}^{A}$, $\mathcal{H}^{B}$, and $\mathcal{H}^{C}$.
		We represent quantum states of the system $A$ as a density operator $\rho^{A}$,
		which is a positive semi-definite operator with unit trace.
		%
		We model our lack of access to a quantum system with the partial trace operation.
		Given a state $\rho^{AB}$ shared between Alice and Bob, we
		can describe Alice's state with the reduced density operator 
		$\rho^{A}=\text{Tr}_{B}\left\{  \rho^{AB}\right\}$,
		where Tr$_{B}$ denotes a partial trace over Bob's degrees of freedom. 
		Let $H(A)_{\rho}\equiv-\text{Tr}\left\{  \rho^{A}\log\rho^{A}\right\}$ 
		denote the von Neumann entropy of the state $\rho^{A}$. 
		%
		A noiseless quantum operation is represented by a unitary operator $U$ which acts on a state 
		$\rho$ by conjugation $U\rho U^\dag$, which we denote as $U \cdot \rho \equiv U\rho U^\dag$.
		Noisy quantum operations are represented by completely positive trace-preserving (CPTP) 
		maps $\mathcal{N}^{A^{\prime}\rightarrow B}$, which accept input states in
		$A^{\prime}$ and produce output states in $B$. 
		%
		%
		%
		%
		Let
		$\operatorname{conv}(\mathcal{R})$ denote the convex closure of any
		geometrical region $\mathcal{R}$. 
		Throughout this paper, logarithms and
		exponents are taken base two unless otherwise specified.

	\subsubsection{Definitions}

		The classical discrete memoryless interference channel (IC) is described by a triple 
		$(\mathcal{X}_1\!\times\!\mathcal{X}_2, p(y_1,y_2|x_1,x_2), \mathcal{Y}_1\!\times\!\mathcal{Y}_2)$,
		where $\mathcal{X}_i$ is a finite set of possible input symbols for Sender $i$ and $\mathcal{Y}_j$ is
		the set of possible output symbols for Receiver $j$.
		%
		

		If we extend this definition to allow both inputs and outputs to be quantum systems we obtain the following:
		
		    \begin{definition}
		                A two party \emph{quantum interference channel} is a triple 
		                $( \mathcal{H}^{A'_1} \otimes \mathcal{H}^{A'_2} , \mathcal{N}^{A^{\prime}_1 A^{\prime}_2 \to B_1B_2}, 
		                 \mathcal{H}^{B_1} \otimes \mathcal{H}^{B_2})$,
		                 where $A'_1$ and $A'_2$ are the two quantum systems that are input to the channel by the senders,
		                 $B_1$ and $B_2$ are the output systems,
		                and $\mathcal{N}^{A^{\prime}_1A^{\prime}_2 \to B_1B_2}$ is a
		                completely positive trace-preserving (CPTP) map.
		    \end{definition}      

		A simpler channel is the \emph{classical-quantum} \mbox{(\textit{c-q})} interference channel,
		where only the outputs are quantum.
		
			\begin{definition}
			            A two party \emph{cc-qq interference channel} is a triple 
			            $(\cX_1 \times \cX_2, \mathcal{N}^{X_1X_2 \to B_1B_2}\!\left(x_1,x_2\right) \equiv \rho^{B_1B_2}_{x_1,x_2},
			              \mathcal{H}^{B_1} \otimes \mathcal{H}^{B_2})$,
			            which models a  general communication network with two classical inputs and
			            a quantum state $\rho^{B_1B_2}_{x_1,x_2}$ as output.
			\end{definition}
			
		In this paper, we focus our attention on the class of classical-quantum interference channels,
		though generalizations of our results to channels with quantum inputs are straightforward.
		We fully specify a \textit{cc-qq} interference channel by the set of output states it produces 
		$\left\{\rho^{B_1B_2}_{x_1,x_2}\right\}_{x_1\in \cX_1, x_2 \in \cX_2}$.
		A classical interference channel with transition probability $p(y_1,y_2|x_1,x_2)$ is a special case of 
		the \textit{cc-qq} channel where the output states are of the form
		$\rho^{B_1B_2}_{x_1,x_2} = \sum_{y_1, y_2} p(y_1,y_2|x_1,x_2) \proj{y_1}^{B_1}\!\otimes\!\proj{y_2}^{B_2}$ where
		$\{\ket{y_1}\}$ and $\{\ket{y_2}\}$ are orthonormal bases of $\cH^{B_1}$ and $\cH^{B_2}$.
	

	\subsubsection{Information processing task}

		The task of communication over an interference channel can be described as follows.
		Using $n$ independent uses of the
		channel, the objective is for Sender~1 to communicate with Receiver~1 at a rate $R_1$
		  and for Sender~2 to communicate with Receiver~2 at a rate $R_2$.
		%
		More specifically, Sender~1 chooses a message $m_1$ from a message set
		$\mathcal{M}_1\equiv \left\{  1,2,\ldots,|\mathcal{M}_1|\right\} $ where $|\mathcal{M}_1|=2^{nR_{1}}$, 
		and Sender~2 similarly chooses a message $m_2$ from a message set 
		$\mathcal{M}_2 \equiv \left\{  1,2,\ldots,|\mathcal{M}_2|\right\}  $ 
		where $|\mathcal{M}_2|=2^{nR_{2}}$. 
		Senders~1 and 2 encode their messages as codewords $x_1^{n}\!\left(  m_1\right)\in \mathcal{X}_1^n$
		and $x_2^{n}\!\left(  m_2\right) \in \mathcal{X}_2^n$ respectively,
		which are then input to the channel.
		%
		The output of the channel is an $n$-fold tensor product state of the form:
		\be
			\mathcal{N}^{\otimes n}\!\left( x_1^{n}(m_1), x_2^{n}(m_2) \right)
			\equiv			
			\rho_{x_2^{n}\left(  m_1\right),  x_2^{n}\left(  m_2\right)  }^{B_{1}^{n}B_{2}^{n}} \ \  \in  \mathcal{H}^{B_1^n B_2^n}.
		\ee



		\noindent
		To decode the message $m_1$ intended for him,
		Receiver~1 performs a positive operator-valued measure (POVM) 
		$\left\{ \Lambda_{m_1}\right\}  _{m_1\in\left\{  1,\ldots,|\mathcal{M}_1|\right\}  }$ on
		the system $B_{1}^n$, the output of which we denote $M^{\prime}_1$. 
		For all $m_1$, $\Lambda_{m_1}$ is a positive operator and $\sum_{m_1}\Lambda_{m_1}=I$.
		Receiver~2 similarly performs a POVM 
		$\left\{  \Gamma_{m_2}\right\} _{m_2\in\left\{  1,\ldots ,|\mathcal{M}_2|\right\}  }$
		  on the system  $B_{2}^n$,
		and the random variable associated with this outcome is denoted $M^{\prime}_2$.
				

		An error event occurs whenever Receiver~1's measurement outcome is different from the message sent by Sender~1
		($M^{\prime}_1 \neq m_1$) or Receiver~2's measurement outcome is different from the message sent by Sender~2 ($M^{\prime}_2 \neq m_2$).
		%
		The overall probability of error for message pair $(m_1,m_2)$ is
		\begin{align*}
			p_{e}\!\left(  m_1,m_2\right)   
				&  \equiv
				\Pr\left\{  
					(M^{\prime}_1,M^{\prime}_2)\neq(m_1,m_2) 
				\right\} \\
				&  =
				\text{Tr}\!
				\left\{  
					\left(  I-\Lambda_{m_1}\otimes\Gamma_{m_2}\right)
					\rho_{x_2^{n}\left(  m_1\right)  x_2^{n}\left(  m_2\right)  }^{B_{1}^{n}B_{2}^{n}} 
				\right\},
		\end{align*}
		where the measurement operator $\left(  I-\Lambda_{m_1}\otimes\Gamma_{m_2}\right)$ represents
		the complement of the correct decoding outcome.

		    \begin{definition}
			An $(n,R_1,R_2,\epsilon)$ code for the interference channel consists
			of two codebooks 
			$\{x^n_1(m_1)\}_{m_1\in \mathcal{M}_1}$
			and 
			$\{x^n_2(m_2)\}_{m_2\in \mathcal{M}_2}$,
			and two decoding POVMs  
			$\left\{ \Lambda_{m_1}\right\}_{m_1\in \mathcal{M}_1}$ 
			and 
			$\left\{  \Gamma_{m_2}\right\}_{m_2\in \mathcal{M}_2}$,
			such that the average probability of error 
			$\overline{p}_{e}$ is bounded from above by $\epsilon$:%
			\begin{align}
				\overline{p}_{e}  \!
				&  \!\equiv \!
					\frac{1}{|\mathcal{M}_1||\mathcal{M}_2|}\sum_{m_1,m_2}p_{e}\!\left(  m_1,m_2\right) 
				 \leq \epsilon.
			\end{align}

		    \end{definition}

%
		
		A rate pair $\left(  R_{1},R_{2}\right)  $ is \textit{achievable} if there exists
		an $\left(  n,R_{1}-\delta,R_{2}-\delta,\epsilon\right)  $ quantum interference channel code
		for all $\epsilon,\delta>0$ and sufficiently large $n$. The channel's \textit{capacity
		region} 
		  is the closure of the set
		of all achievable rates.

\section{Decoding strategies for quantum multiple~access channels}
\label{sec:MAC}

	The quantum interference channel described by 
	$(\mathcal{X}_1 \times \mathcal{X}_2,  \rho_{x_1,x_2}^{B_1B_2}, \mathcal{H}^{B_1} \otimes \mathcal{H}^{B_2})$
	induces two multiple access (MAC) sub-channels.
	More specifically \MACone is the channel to Receiver~1 given by
	$(\mathcal{X}_1 \times \mathcal{X}_2,  \rho_{x_1,x_2}^{B_1}= 
	\mathop{\textrm{Tr}}_{B_2}\!\!\left\{\rho_{x_1,x_2}^{B_1B_2}\right\}, \mathcal{H}^{B_1} )$,
	and \MACtwo 
	is the channel to Receiver~2 defined by $(\mathcal{X}_1 \times \mathcal{X}_2,  \rho_{x_1,x_2}^{B_2}, \mathcal{H}^{B_2} )$.
	In order to better understand the interference channel problem we first consider
	the different decoding strategies for the individual receivers.
	%
	In this section 
	we analyze three types of decoding strategies for quantum multiple access channels,
	and then in Section \ref{sec:QIC} we use each of these to build a corresponding interference channel code.

	Winter found a single-letter formula for the capacity of the classical-quantum 
	multiple access channel \cite{winter2001capacity}. 

	\begin{theorem}[Theorem 10 in  \cite{winter2001capacity}]
		\label{thm:cqmac-capacity}
		The capacity region for the classical-quantum multiple
		access channel $(\mathcal{X}_1 \times \mathcal{X}_2,  \rho_{x_1,x_2}^{B}, \mathcal{H}^{B} )$
		is given by 
		\begin{equation}
	        		\nonumber
	        		 \mathcal{C}_{\textrm{MAC}}  
				\equiv  
					\operatorname{conv}
					\bigcup_{ p_{X_{1}}\!, p_{X_{2}}\! }
					\!\!\!\!\! \{ (R_1,R_2) \in \bbR^2 | \text{ \emph{Eqns.} \eqref{winterThmEqnsOne}-\eqref{winterThmEqnsThree} } \} 
		\end{equation}
		\vspace{-5mm}
	        \bea 
	            R_1             &\leq&    I(X_1;B|X_2)_\theta,  \label{winterThmEqnsOne} \\
	            R_2             &\leq&    I(X_2;B|X_1)_\theta,  \label{winterThmEqnsTwo} \\
	            R_1+R_2   &\leq&    I(X_1X_2;B)_\theta, \label{winterThmEqnsThree}
	        \eea
	        where the information quantities are taken with respect to the 
	        classical-quantum state
	        $\theta^{X_{1}X_{2}B}$ given by 
		\be
		       	\sum_{x_1,x_2} 
				p_{X_{1}}\!\left(x_{1}\right)p_{X_{2}}\!\left( x_{2}\right)
				\ketbra{x_1}{x_1\!}^{X_1} \!\!
				\otimes \!
				\ketbra{x_2}{x_2}^{X_2} \!
				\otimes \!
				\rho^{B}_{x_1x_2}.
		\ee	        	        
		%

	\end{theorem}	




	\subsection{Successive decoding}
	\label{sec:mac-succ-decoding}

	
			%
			The technique used by Winter to prove the achievability of the rates  in Theorem~\ref{thm:cqmac-capacity}
			is called \emph{successive decoding}.
			%
			For a given pair of probability distributions
			$p \equiv p_{X_{1}},p_{X_{2}}$,
			the achievable rate region 
			has the form of a pentagon bounded by the three inequalities 
			in equations \eqref{winterThmEqnsOne}-\eqref{winterThmEqnsThree} and two rate positivity conditions.
			The two dominant vertices of this rate region have coordinates 
			$\alpha_{p}
			\equiv (I(X_1;B)_\theta, I(X_2;B|X_1)_\theta)$ and 
			$\beta_{p}
			\equiv (I(X_1;B|X_2)_\theta, I(X_2;B)_\theta)$
			and correspond to two alternate successive decoding strategies.

			To achieve the rates of  $\alpha_{p}$,  
			the receiver  first performs a measurement 
			$\big\{  \Lambda^\alpha_{m_1}\big\}$ to decode the 
			message $m_1$,  
			and then performs a second measurement 
			to recover the message $m_2$.
			%
			The second measurement 
			is $\big\{  \Lambda^\alpha_{m_2|m_1}\big\}$, where we have
			indicated that the second measurement is conditional on  $m_1$.
			By using these POVMs, Winter shows that if the rates $(R_1,R_2)$ satisfy
			\begin{align}
				R_{1}  &  \leq I\left(  X_1;B\right)_{\theta},			\label{first-rate}\\
				R_{2}  &  \leq I\left(  X_2;B|X_1\right)_{\theta},		\label{second-rate}
			\end{align}
			then the expected success probability 
			asymptotically approaches one:
			\begin{align}
			& \mathop{\mathbb{E}}_{X_1^{n},X_2^{n}}
				\left\{ 
					\frac{1}{|\mathcal{M}_1||\mathcal{M}_2|}
					\sum_{m_1,m_2} 
					\text{Tr}
						\left\{
							\Lambda^w_{m_1,m_2}
							\rho_{X_1^{n}\left(  m_1\right)  ,X_2^{n}\left(	m_2\right) }
						\right\}  
				\right\} \ \ \ \ \nonumber \\
			& \qquad  \qquad \qquad \qquad \qquad \qquad \qquad \qquad 
			\geq1-\epsilon,
			\end{align}
			where we have informally denoted by $\Lambda^w_{m_1,m_2}$
			the successive measurements of $\Lambda^\alpha_{m_1}$ followed by $\Lambda^\alpha_{m_2|m_1}$.

			The rate point $\beta_{p}$ corresponds to the alternate decode ordering
			where the receiver decodes the message $m_2$ first and $m_1$ second.
			The corner points $\alpha_{p}$ and 	$\beta_{p}$ 
			are important because given codes
			that achieve them, we can use  \emph{time-sharing} and 
			\emph{resource wasting} to obtain all other rate pairs in the region.
			The $M$-sender MAC has $M!$ such corner points, one for each permutation
			of the decode ordering.

	\subsection{Quantum simultaneous decoding}

		Another approach for achieving the capacity of the multiple access channel, which does not use time-sharing,
		is simultaneous decoding.
		The analysis of the \emph{classical} simultaneous decoder
		is a straightforward application of the joint typicality lemma to bound the probability 
		of the different decoding error events that may occur \cite{el2010lecture}.
		In the quantum case, we can similarly identify four different error events,
		but the construction of a measurement operator based on typical subspace projectors
		is more difficult to analyse because the different typical projectors may not commute in general.
		
		In this section we prove that a quantum simultaneous decoder exists for multiple 
		access channels with two senders and formulate Conjecture~\ref{conj:sim-dec} regarding the 
		existence of a simultaneous decoder for three-sender multiple access channels.

		\begin{theorem}[Two-sender simultaneous decoding]
		\label{thm:sim-dec-two-sender}
			Consider  
			the \emph{cc-q} multiple access channel
			with two senders and a single receiver $(\mathcal{X}_1 \times \mathcal{X}_2,  \rho_{x_1,x_2}^{B}, \mathcal{H}^{B} )$.
			Let $p_{X_i}$ be a distribution on $\cX_i$ and $\cM_i \equiv \{1, \dots, 2^{n(R_i-\delta)}\}$
			for $i \in \{1,2\}$ and $\delta>0$.
			Define the random codebooks $\{X_1^n(m_1)\}_{m_1 \in \cM_1}$ and $\{X_2^n(m_2)\}_{m_2 \in \cM_2}$ 
			generated from the product distributions $p_{X_1^n}$ and $p_{X_2^n}$ respectively. 
			There exists a simultaneous decoding POVM 
			$\left\{  \Lambda_{m_1,m_2}\right\}_{m_1\in \mathcal{M}_1,m_2\in\mathcal{M}_2}$,
			with  expected average probability of error bounded from above by $\epsilon$
			for all $\epsilon,\delta>0$ and sufficiently large $n$
			provided the rates $R_1,R_2$ satisfy inequalities \eqref{winterThmEqnsOne}-\eqref{winterThmEqnsThree}.
		\end{theorem}

		The proof proceeds by random coding arguments using the
		properties of projectors onto the typical subspaces of the output states
		\cite{wilde2011book} and the \emph{square-root} measurement.
		Note that Sen proved the same result using different techniques in \cite{S11a}.
		\ifthenelse{\boolean{WITHAPDX}}{
		        See Appendix~\ref{sec:typ-review} for a review of the properties
		        of typical subspaces.
		    }
		    {
		    }
		%
	
		\begin{proof}
		%
		%
		Let state $\rho_{m_1,m_2}\equiv \rho_{x_1^n(m_1)  ,x_2^{n}\left(  m_2\right)  }$ denote
		the output of the $n$ uses of the channel when codewords $x_1^n(m_1) $ and
		$x_2^n(m_2)  $ are input. 
		Let $\PIonetwo \equiv \Pi_{\rho_{x_1^n(m_1)  ,x_2^{n}\left(  m_2\right)
		},\delta}^{n}$ be the conditionally typical projector for that state.
		Consider the following code-averaged output states:
		\begin{align}
			\bar{\rho}_{x_1} &  \equiv
				\sum_{x_2}p_{X_2}\!\left(  x_2\right)  \rho_{x_1,x_2},\label{eq:rho_x} \\
			\bar{\rho}_{x_2} &  \equiv
				\sum_{x_1}p_{X_1}\!\left(  x_1\right)  \rho_{x_1,x_2},\label{eq:rho_y} \\
			\bar{\rho} &  \equiv
				\sum_{x_1,x_2}p_{X_1}\!\left(  x_1\right)  p_{X_2}\!\left(  x_2\right) \rho_{x_1,x_2}.\label{eq:rho}
		\end{align}		
		Let $\PIone \equiv \Pi_{\bar{\rho}_{x_1^n(m_1)  },\delta}^{n}$ be the
		conditionally typical projector for the tensor product state 
		$\bar{\rho}_{m_1} \equiv \bar{\rho}_{x_1^n(m_1)  }$ defined by (\ref{eq:rho_x}) for $n$ uses of the
		channel.
		Let $\PItwo \equiv \Pi_{\bar{\rho}_{x_2^{n}\left(  m_2\right)  },\delta}^{n}$ 
		be the conditionally typical projector for the tensor product state 
		$\bar{\rho}_{m_2} \equiv  \bar{\rho}_{x_2^{n}\left(  m_2\right)  }$ 
		defined by (\ref{eq:rho_y}) 
		and finally let $\Pi_{\bar{\rho},\delta}^{n}$ be the typical
		projector for the 
		state $\bar{\rho}^{\otimes n}$ defined by~(\ref{eq:rho}).

		The detection POVM\ $\left\{  \Lambda_{m_1,m_2}\right\}  $ has the following form:
		\vspace{-2mm}
		\begin{align}
		\Lambda_{m_1,m_2} &  \!\! \equiv \!
		\left(  
			\sum_{m_1^{\prime},m_2^{\prime}}\Pi_{m_1^{\prime},m_2^{\prime}}^{\prime}
		\right)^{\!\!\!-\frac{1}{2}}
		\!\! \Pi_{m_1,m_2}^{\prime}
		\left(
			\sum_{m_1^{\prime},m_2^{\prime}}\Pi_{m_1^{\prime},m_2^{\prime}}^{\prime}
		\right)^{\!\!-\frac{1}{2}}\!\!, 
		\nonumber
		\end{align}
		where 
		\begin{align}
		%
		\Pi_{m_1,m_2}^{\prime} &  \equiv 
		\Pi_{\bar{\rho},\delta}^{n} \cdot \PIone \cdot \PIonetwo  , \label{eq:proj-sandwitch} 
		\end{align}
		is a positive operator which consists of three typical projectors ``sandwiched'' together.



		The average error probability of the code is given by:%
		\begin{equation}
			\overline{p}_{e}
			\equiv
			\frac{1}{|\mathcal{M}_1||\mathcal{M}_2|}\sum_{m_1,m_2}
				\text{Tr}\left\{  
					\left(I-\Lambda_{m_1,m_2}\right)  
					\rho_{m_1,m_2}
				\right\}.
			\label{eq:avg-error-prob}%
		\end{equation}
		
		One key insight for the proof 
		is the substitution of the output state 
		$\rho_{m_1,m_2}$ 
		with a smoothed version:
		\be
			\tilde{\rho}_{m_1,m_2} \equiv \PItwo \rho_{m_1,m_2} \PItwo,
			\label{smoothed-rho}
		\ee
		and bounding \eqref{eq:avg-error-prob} from above as follows:
		\begin{align}
			\overline{p}_{e}
			\leq &
			\frac{1}{|\mathcal{M}_1||\mathcal{M}_2|}\!\sum_{m_1,m_2}\!\!\Bigg[  
				\text{Tr}\left\{  
					\left(I-\Lambda_{m_1,m_2}\right)  
					\tilde{\rho}_{m_1,m_2}
				\right\} \hspace{8mm}
				\nonumber \\[-5mm]
			 & \qquad \qquad  \qquad \qquad \quad
			 	+ \Vert \tilde{\rho}_{m_1,m_2} - \rho_{m_1,m_2} \Vert_1
			\Bigg].
			\label{eq:error-bound-pre-HN}%
		\end{align}
		To obtain \eqref{eq:error-bound-pre-HN}, we used the inequality 
		\be
		\text{Tr}\left\{  \Lambda\rho\right\}  \leq\text{Tr}\left\{ \Lambda
		\sigma\right\}  +\left\Vert \rho-\sigma\right\Vert _{1},
		\label{eqn:tr-trick}
		\ee
		which holds for all  operators such that   $0\leq \rho, \sigma, \Lambda \leq I$.

		The Hayashi-Nagaoka operator inequality applies to all positive operators $T$ and 
		$S$ where $0\leq S\leq I$ \cite{hayashi2003general}:%
		\[
			I-\left(  S+T\right)^{-\frac{1}{2}}S\left(  S+T\right)  ^{-\frac{1}{2}}
			\leq
			2\left(  I-S\right)  +4T.
		\]
		Choosing $S=\Pi_{m_1,m_2}^{\prime}, \ T=\sum_{\left(  m_1^{\prime},m_2^{\prime}\right)
		\neq\left(  m_1,m_2\right)  }\Pi_{m_1^{\prime},m_2^{\prime}}^{\prime}$, 
		we apply the above operator inequality to bound the average error
		probability of the first term in  (\ref{eq:error-bound-pre-HN}) as:
		\begin{align}
			\overline{p}_{e}
			\leq &
			\frac{1}{|\mathcal{M}_1||\mathcal{M}_2|}\!\sum_{m_1,m_2}\!\!\Bigg[  
					2\text{Tr}\left\{  
					\left(I-\Pi_{m_1,m_2}^{\prime}\right)  \tilde{\rho}_{m_1,m_2} 
					\right\} 			\label{eq:error-bound-HN}  \\
			 &  \!\!\!
			 	+4\hspace{-0.9cm}\sum_{\left(  m_1^{\prime},m_2^{\prime}\right)  \neq\left(m_1,m_2\right)  } 
				\hspace{-0.8cm} \text{Tr}\left\{  
					\Pi_{m_1^{\prime},m_2^{\prime}}^{\prime}\tilde{\rho}_{m_1,m_2} 
					\right\}  
					+ 
						\Vert \tilde{\rho}_{m_1,m_2} \!- \!\rho_{m_1,m_2} \Vert_1
			\Bigg].   \nonumber
		\end{align}
		%

		We apply a random coding argument to bound the expectation of 
		the average error probability in \eqref{eq:error-bound-HN}.
		%
		A bound on the first term follows from the following argument:
		\begin{align}
		&  \mathop{\mathbb{E}}_{X^n_1,X^n_2} \text{Tr}\left\{  \Pi_{m_1,m_2}^{\prime}\tilde{\rho}_{m_1,m_2} \right\}  \hspace{4.5cm} \nonumber \\
		&\quad  =\mathop{\mathbb{E}}_{X^n_1,X^n_2} \text{Tr}\left\{  
		\Pi_{\bar{\rho},\delta}^{n} \cdot \PIone \cdot \PIonetwo
		\  \PItwo  \rho_{m_1,m_2} \PItwo
		\right\}  \nonumber\\
		&  \quad  \geq \mathop{\mathbb{E}}_{X^n_1,X^n_2} \text{Tr}\left\{  
		\PIonetwo \rho_{m_1,m_2} \right\}  \nonumber \\
		& \qquad \qquad \ \ \ 
		-\mathop{\mathbb{E}}_{X^n_1,X^n_2}
		\left\Vert 
		\PItwo \rho_{m_1,m_2 } \PItwo  -\rho_{m_1,m_2} 
		\right\Vert _{1}\nonumber\\
		& \qquad \qquad \ \ \  
		-\mathop{\mathbb{E}}_{X^n_1,X^n_2} 
		\left\Vert 
		\PIavg  \rho_{m_1,m_2}  \PIavg  -\rho_{m_1,m_2}  \right\Vert _{1}\nonumber\\
		& \qquad \qquad \ \ \ 
		-\mathop{\mathbb{E}}_{X^n_1,X^n_2}
		\left\Vert 
		\PIone \rho_{m_1,m_2} \PIone-\rho_{m_1,m_2}
		\right\Vert _{1}\nonumber\\
		& \quad   \geq1-\epsilon-6\sqrt{\epsilon}. \label{eq:first-error-chain}%
		\end{align}
		%
		The first inequality follows from \eqref{eqn:tr-trick} applied three times.
		The second inequality follows from the \emph{Gentle Measurement Lemma} for ensembles 
		\cite[Lemma 9.4.3]{wilde2011book} 
		and the properties of entropy-typical projectors \cite[Section 14.2.2]{wilde2011book}.
		The same reasoning is used to obtain a bound the expectation of the smoothing-penalty term 
		in equation~\eqref{eq:error-bound-HN}:  $\mathbb{E}_{X^n_1,X^n_2}  \Vert \tilde{\rho}_{m_1,m_2} \!- \!\rho_{m_1,m_2} \Vert_1 \leq  2\sqrt{\epsilon}$.


		We decompose the second term in \eqref{eq:error-bound-HN} into 
		three error events, each representing a different type of decoding error:
		\vspace{-2mm}
		\begin{align}
		&\hspace{-4mm} 
		\sum_{\left(  m_1^{\prime},m_2^{\prime}\right)  \neq\left(  m_1,m_2\right)  }
		\hspace{-9mm}
		\text{Tr}\left\{  \Pi_{m_1^{\prime},m_2^{\prime}}^{\prime}
		\tilde{\rho}_{m_1,m_2}
		\right\}  
		= \nonumber \\[-2mm]
		& \qquad \qquad  =
			\sum_{m_1^{\prime}\neq m_1}\text{Tr}\left\{  \Pi_{m_1^{\prime},m_2}^{\prime}
			\tilde{\rho}_{m_1,m_2}
			\right\}  
			\tag{E1} \label{eq:err-one}\\[-1mm]
		& \qquad \qquad \qquad 
			+\sum_{m_2^{\prime}\neq m_2}\text{Tr}\left\{\Pi_{m_1,m_2^{\prime}}^{\prime}
			\tilde{\rho}_{m_1,m_2}
			\right\} \tag{E2} \label{eq:err-two} \\[-1mm]
		& \qquad \qquad \qquad 
			+ \hspace{-6mm} \sum_{m_1^{\prime}\neq m_1,m_2^{\prime}\neq m_2}
			\hspace{-6mm} \text{Tr}\left\{  \Pi_{m_1^{\prime
			},m_2^{\prime}}^{\prime}\tilde{\rho}_{m_1,m_2}
			\right\}. \tag{E12}
			\label{eq:err-both}%
		\end{align}%
		%
		\vspace{-1mm}%
		The expectation the over random choice of codebook for event \eqref{eq:err-one},
		the event that $m_1$ is decoded incorrectly, is as follows:
		\vspace{-1mm}
		\begin{align}
		 \mathop{\mathbb{E}}_{X^n_1,X^n_2}&
		\big\{  \sum_{m_1^{\prime}\neq m_1}
		\text{Tr}\left[  \Pi_{m_1^{\prime},m_2}^{\prime}
		\tilde{\rho}_{m_1,m_2}
		\right]  \big\}  \nonumber\\[-2mm]
		&  = \!\!\!\sum_{m_1^{\prime}\neq m_1}\mathop{\mathbb{E}}_{X^n_2}\left\{  \text{Tr}\left[
		\mathop{\mathbb{E}}_{X^n_1}\left\{  
		\Pi_{m_1^{\prime},m_2}^{\prime} \right\}  
		\mathop{\mathbb{E}}_{X^n_1}\left\{
		  \tilde{\rho}_{m_1,m_2}
		  \right\} 
		 \right]
		\right\}  \nonumber\\
		&  = \!\!\!\sum_{m_1^{\prime}\neq m_1}\mathop{\mathbb{E}}_{X^n_1X^n_2}\left\{  \text{Tr}\left[
			\Pi_{m_1^{\prime},m_2}^{\prime} 
		 	\PItwo
			\bar{\rho}_{m_2}
			\PItwo \right]
		\right\}  \nonumber \\
		&  \leq 2^{-n\left[  H\left(B|X_2\right) -\delta\right]}  
		\!\! \sum_{m_1^{\prime}\neq m_1}\mathop{\mathbb{E}}_{X^n_1X^n_2}\left\{  \text{Tr}\left[
		\Pi_{m_1^{\prime},m_2}^{\prime} 
		 	\PItwo \right]
		\right\}  \nonumber 
		\end{align}%
		\vspace{-1mm}%
		The first equality follows because the codewords labeled by $m_1^{\prime}$ and $m_1$ are independent.
		The second equality comes from the definition of the averaged code state 
		$\bar{\rho}_{m_2} \equiv  \bar{\rho}_{x_2^{n}\left(  m_2\right)  }$.
		The last inequality follows from:
		\[
		\PItwo \bar{\rho}_{m_2} \PItwo 
		\leq
		2^{-n\left[  H\left(B|X_2\right) -\delta\right]  }
		\PItwo.
		\]

		We focus our attention on the expression inside the trace:
		\begin{align}
		\!\!\!\!\!\text{Tr}&\left[
		\Pi_{m_1^{\prime},m_2}^{\prime}  \ 
		 	\PItwo \right] \nonumber \\
		& = 
		\text{Tr}\left[
		\Pi_{\bar{\rho},\delta}^{n}  \cdot \PIonepr \cdot \PIoneprtwo
		 \ \  \PItwo \right]
		 \nonumber \\
		& = 
		\text{Tr}\left[
		\Pi_{\bar{\rho},\delta}^{n}  \
		\PIonepr \
		\PIoneprtwo  \
		\PIonepr \
		\Pi_{\bar{\rho},\delta}^{n} 
		\ \ \PItwo		
		 \right] \nonumber \\
		& = 
		\text{Tr}\left[
		\PIonepr \
		\Pi_{\bar{\rho},\delta}^{n} \
		\PItwo		\ 
		\Pi_{\bar{\rho},\delta}^{n}  \
		\PIonepr  \
		\PIoneprtwo
		 \right] \nonumber \\		
		& \leq 
		\text{Tr}\left[
		\PIoneprtwo 
		 \right]. \nonumber 
		\end{align}
		In the first step we substituted the definition of $\Pi_{m_1,m_2}^{\prime}$ from
		equation \eqref{eq:proj-sandwitch}.
		The rest of the equalities follow from the cyclicity of trace.
		%
		The inequality follows from
		\vspace*{-1mm}
		\begin{align}
		\PIonepr \PIavg \PItwo \PIavg \PIonepr 
		 \leq 
		 \PIonepr \PIavg \PIonepr
		 \leq 
		 \PIonepr
		 \leq
		 I. \label{positive-op-anihilation}
		\end{align}

		Continuing, we obtain the following bound on the expected probability 
		of error event \eqref{eq:err-one}:
		\vspace{-2mm}
		\begin{align}
		\!\!\mathop{\mathbb{E}}_{X^n_1,X^n_2}
		\!\!\!\!\!\!\left\{ \eqref{eq:err-one}
		\right\}  
		& \leq
		2^{-n\left[  H\left(  B|X_2\right)-\delta\right]  } 
		\!\!\!\!
		\sum_{m_1^{\prime}\neq m_1}
		\mathop{\mathbb{E}}_{X^n_1,X^n_2}
		\left\{  \text{Tr}\left\{  \PIoneprtwo 
		\right\}  \right\}  \nonumber\\
		&  \leq
		2^{-n\left[  H\left(  B|X_2\right)-\delta\right]  }
		\sum_{m_1^{\prime}\neq m_1}
		2^{n\left[  H\left(  B|X_1X_2\right)+\delta\right]  }\nonumber\\
		&  \leq
		|\mcal{M}_1|\  2^{-n\left[  I\left(X_1;B|X_2\right)  -2\delta\right]  }.
		\label{eq:err-one-bound}
		\end{align}
		The second inequality in \eqref{eq:err-one-bound} follows from the bound%
		\begin{align}
		\text{Tr}\{
		\PIonetwo
		\}\leq2^{n\left[  H\left(  B|X_1X_2\right)  +\delta\right]  }%
		\nonumber
		\end{align}
		on the rank of a conditionally typical projector.
		
		We employ a different argument to bound the probability of the second error event 
		\eqref{eq:err-two} based on the following fact
		\begin{align}
			\PIonetwo 
			& \leq 
				2^{n[H(B|X_1X_2) + \delta] } 
				\PIonetwo
				\rho^B_{m_1,m_2}
				\PIonetwo \nonumber \\
			& =
				2^{n[H(B|X_1X_2) + \delta] } 
				\sqrt{\rho^B_{m_1,m_2}}
				\PIonetwo
				\sqrt{\rho^B_{m_1,m_2}}
				\nonumber \\
			& \leq
				2^{n[H(B|X_1X_2) + \delta] } 
				\rho^B_{m_1,m_2},
		\end{align}
		which we refer to as the \emph{projector trick} \cite{GLM10}.
		The first inequality is the standard lower bound on the eigenvalues of
		$\rho^B_{m_1,m_2}$ expressed as an operator upper bound on the projector $\PIonetwo$.
		The equality follows because the state and its typical projector commute.
		The last inequality follows from $0 \leq \PIonetwo \leq I$.
		
		Continuing, 
		\vspace{-3mm}
		\begin{align}
		& 	\hspace{-3mm}
			\mathop{\mathbb{E}}_{X^n_1,X^n_2}
			\!\!
			\Big\{ \eqref{eq:err-two}
			\Big\} = 
			\mathop{\mathbb{E}}_{X^n_1,X^n_2}
		\!\!
		\left\{ 
		\sum_{m_2^{\prime}\neq m_2}\text{Tr}\left[\Pi_{m_1,m_2^{\prime}}^{\prime}
			\tilde{\rho}_{m_1,m_2}
			\right] 
		\right\} \nonumber \\
		%
		&  = \!\!\!\sum_{m_2^{\prime}\neq m_2}\mathop{\mathbb{E}}_{X^n_1}\left\{  \text{Tr}\left[
		\mathop{\mathbb{E}}_{X^n_2}\left\{  
		\Pi_{m_1,m_2^{\prime}}^{\prime} \right\}  
		\mathop{\mathbb{E}}_{X^n_2}\left\{
		  \tilde{\rho}_{m_1,m_2}
		  \right\} 
		 \right]
		\right\}  \nonumber\\
		&  = \!\!\!\sum_{m_2^{\prime}\neq m_2}
		\!\!\!\mathop{\mathbb{E}}_{X^n_1}\left\{  \text{Tr}\left[
		\mathop{\mathbb{E}}_{X^n_2}\left\{  
			\Pi_{\bar{\rho},\delta}^{n}  \cdot \PIone \cdot \PIonetwopr
		\right\}  
		\mathop{\mathbb{E}}_{X^n_2}\left\{
		  \tilde{\rho}_{m_1,m_2}
		  \right\} 
		 \right]
		\right\}  \nonumber\\
		&  = \!\!\!\sum_{m_2^{\prime}\neq m_2}
		\!\!\!\mathop{\mathbb{E}}_{X^n_1}\!\!\left\{  
		\!\!\text{Tr}\!\!\left[
		\Pi_{\bar{\rho},\delta}^{n}\!
		\mathop{\mathbb{E}}_{X^n_2}\!\!\left\{  
			  \PIone\PIonetwopr\PIone
		\right\}\!
		\Pi_{\bar{\rho},\delta}^{n}
		\!\!\mathop{\mathbb{E}}_{X^n_2}\!\!\left\{
		  \tilde{\rho}_{m_1,m_2}
		  \right\} 
		 \right]
		\!\right\}  \nonumber
		\end{align}
		We focus our attention on the first expectation term:
		\begin{align}
		&
		\!\!\!\!\!\!\!\!\!
		\mathop{\mathbb{E}}_{X^n_2}\!\!\left\{  
			  \PIone\PIonetwopr\PIone
		\right\}  		  \nonumber \\
		& \ \ 
		\leq 2^{n[H(B|X_1X_2) + \delta] } 
		\mathop{\mathbb{E}}_{X^n_2}\!\!\left\{  
			  \PIone \rho^B_{m_1,m_2^{\prime}} \PIone
		\right\}  		  \nonumber \\
		& \ \ 
		= 2^{n[H(B|X_1X_2) + \delta] } 
		\PIone
		\mathop{\mathbb{E}}_{X^n_2}\!\!\left\{  
			   \rho^B_{m_1,m_2^{\prime}} 
		\right\} 
		\PIone
		\nonumber \\
		& \ \ 
		= 2^{n[H(B|X_1X_2) + \delta] } 
		\PIone
		\bar{\rho}_{m_1}
		\PIone
		\nonumber \\
		& \ \ 
		\leq 2^{n[H(B|X_1X_2) + \delta] } 
		2^{-n[H(B|X_1) - \delta] } 
		\PIone
		\nonumber \\
		& \ \ 
		= 2^{-n[I(X_2;B|X_1)- 2\delta] } 
		\PIone
		\nonumber
		\end{align}

		Substituting back into the expression for the error bound,
		we obtain:
		\begin{align}
		& 	\hspace{-3mm}
			\mathop{\mathbb{E}}_{X^n_1,X^n_2}
			\!\!\!
			\{ \eqref{eq:err-two}
			\}  
		 \leq
		2^{-n[I(X_2;B|X_1)- 2\delta] } 
		\!\!\!\!\sum_{m_2^{\prime}\neq m_2} \!\!
		\!\!\text{Tr}\!\left[
		\Pi_{\bar{\rho},\delta}^{n}\!
			  \PIone
		\Pi_{\bar{\rho},\delta}^{n}
		  \tilde{\rho}_{m_1,m_2}
		 \right]
		\nonumber \\
		&  =
		2^{-n[I(X_2;B|X_1)- 2\delta] } 
		\!\!\!\sum_{m_2^{\prime}\neq m_2}
		\!\!\text{Tr}\!\left[
		\Pi_{\bar{\rho},\delta}^{n}\!
			  \PIone
		\Pi_{\bar{\rho},\delta}^{n}
		 \PItwo \rho_{m_1,m_2} \PItwo
		 \right]
		\nonumber \\
		&  =
		2^{-n[I(X_2;B|X_1)- 2\delta] } 
		\!\!\!\sum_{m_2^{\prime}\neq m_2}
		\!\!\text{Tr}\!\left[
		\PItwo
		\Pi_{\bar{\rho},\delta}^{n}\!
			  \PIone
		\Pi_{\bar{\rho},\delta}^{n}
		 \PItwo \rho_{m_1,m_2} 
		 \right]
		\nonumber \\
		&  \leq
		2^{-n[I(X_2;B|X_1)- 2\delta] } 
		\!\!\!\sum_{m_2^{\prime}\neq m_2}
		\!\!\text{Tr}\!\left[
		 \rho_{m_1,m_2} 
		 \right]
		\nonumber \\ 
		& 
		 \leq
		2^{-n\left[  I\left(  X_2;B|X_1\right)  -2\delta\right]  }
		|\mathcal{M}_2|,  \label{eq:err-two-bound}
		\end{align}
		\noindent
		The second inequality follows from an argument analogous to
		\eqref{positive-op-anihilation}.

		By a different argument involving averaged states, 
		we bound the probability of the third error event as: %
		\begin{align}
		\mathop{\mathbb{E}}_{X^n_1,X^n_2}
		\!\!
		\Big\{\eqref{eq:err-both}
		\Big\}  
		& \leq
		|\mathcal{M}_1||\mathcal{M}_2|\ 2^{-n\left[  I\left(  X_1X_2;B\right)-2\delta\right]  }.
		\label{eqn:sum-rate-bound}
		\end{align}

		Combining the bounds from equations 
		\eqref{eq:first-error-chain},
		\eqref{eq:err-one-bound},
		\eqref{eq:err-two-bound}, 
		\eqref{eqn:sum-rate-bound} and the smoothing penalty, 
		we get the following bound on the expectation
		of the average error probability:%
		\begin{multline*}
		\mathop{\mathbb{E}}_{X_1^{\prime n},X_2^{\prime n}}\!\!
		\Big\{  \overline{p}_{e}\Big\}
		\leq2\left(  \epsilon+6\sqrt{\epsilon}\right) + 2\sqrt{\epsilon} \\
		+4\bigg[  |\mathcal{M}_1|\ 2^{-n\left[  I\left(
		X_1;B|X_2\right)  -2\delta\right]  }+|\mathcal{M}_2|\ 2^{-n\left[  I\left(  X_2;B|X_1\right) 
		-2\delta\right]  }\\ 
		+ |\mathcal{M}_1||\mathcal{M}_2|\ 2^{-n\left[
		I\left(  X_1X_2;B\right)  -2\delta\right]  }\bigg]  .
		\end{multline*}
		Thus, if we choose the message sets sizes to be 
		$|\mcal{M}_1|  =2^{n\left[  R_{1}-3\delta\right]  }$,
		and 
		$|\mcal{M}_2|   =2^{n\left[  R_{2}-3\delta\right]  }$,
		the expectation of the average error probability vanishes whenever the
		rates $R_{1}$ and $R_{2}$ obey the inequalities:
		\begin{align*}
		R_{1}-\delta   <I\left(  X_1;B|X_2\right), & \ \ \ 
		R_{2}-\delta   <I\left(  X_2;B|X_1\right)  ,\\
		R_{1}+R_{2}-4\delta &  <I\left(  X_1X_2;B\right).
		\end{align*}
		Given that $\delta>0$ is an arbitrarily small number
		the bounds in the statement of the theorem follow.
		\end{proof}


		We now state our conjecture regarding the existence of a quantum simultaneous
		decoder
		for the three-sender case.

		\begin{conjecture}[Three-sender QMAC simultaneous decoding]
			\label{conj:sim-dec}
			Let $\mathcal{C}_{\textrm{3MAC}}$ denote the capacity region 
			of a \emph{ccc-q} multiple access channel with three senders:
			 $x_1,x_2,x_3 \rightarrow\rho^B_{x_1,x_2,x_3}$.
			Let  $\{X_i^n(m_i)\}_{m_i \in \cM_i}$, for $i \in \{1,2,3\}$ 
			be random codebooks generated according to the product distributions $p^n_{X_i^n}$
			with messages sets $\cM_i \equiv \{1, \dots, 2^{n(R_i-\delta)}\}$
			with $\delta>0$.
			%
			There exists a simultaneous decoding POVM 
			$\left\{  \Lambda_{m_1,m_2,m_3}\right\}
			$,
			with  expected average probability of error bounded from above by $\epsilon$
			for all $\epsilon,\delta>0$ and sufficiently large $n$
			for any rate triple $(R_1,R_2,R_3) \in \mathcal{C}_{\textrm{3MAC}}$.
		\end{conjecture}


		Were this conjecture true, it would form
		 a fundamental building block for multiuser information theory.
		Obtaining a proof might allow us to directly adapt 
		many of the known classical techniques
		of classical multiuser information theory to the quantum setting.
		Indeed, many coding theorems in classical network information
		theory exploit a simultaneous decoding approach (jointly typical decoding)
		\cite{el2010lecture}. 
		%
		%

		We can prove that simultaneous decoding works for a special class of three-sender MACs 
		for which the averaged output states (defined analogously to \eqref{eq:rho_x}  and \eqref{eq:rho_y})
		satisfy the following commutation relations:
		$\left[  \bar{\rho}_{x_1,x_2},\bar{\rho}_{x_2,x_3}\right]=0$, 
		$\left[  \bar{\rho}_{x_1,x_3},\bar{\rho}_{x_1,x_2}\right]=0$, 
		$\left[  \bar{\rho}_{x_1,x_3},\bar{\rho}_{x_2,x_3}\right]=0$, 
		$\forall x_1,x_2,x_3$.
		These commutation relations imply that the corresponding typical projectors
		commute and thus give a simpler construction of the measurement operator.
		
		Furthermore, we can prove that a quantum simultaneous decoder exists for a random code 
		provided that the rates $R_1$, $R_2$ and $R_3$
		satisfy a set of stronger constraints involving min-entropies.
		We invite the reader to consult  \cite{FHSSW11}  for further details 
		about these special cases.



\ifthenelse{\boolean{WITHHMINS}}{


		%
		Furthermore, we can prove that a quantum simultaneous decoder exists for a random code 
		provided that the rates $R_1$, $R_2$ and $R_3$
		satisfy a set of stronger constraints involving min-entropies.
		The min-entropy $H_{\min}\!\left(  B\right)_{\rho}$ of a quantum
		state $\rho^{B}$ is equal to the negative logarithm of its maximal eigenvalue: 
		$
		H_{\min}\left(  B\right) _{\rho}
		\equiv
			-\log\left(  \inf_{\lambda\in \mathbb{R}}
					\left\{  \lambda:\rho\leq\lambda I\right\}  
				\right),
		$

		and the conditional min-entropy of a classical-quantum state $\rho^{XB}%
		\equiv\sum_{x}p_{X}\left(  x\right)  \left\vert x\right\rangle \left\langle
		x\right\vert ^{X}\otimes\rho_{x}^{B}$ with classical system $X$ and quantum
		system $B$ is   
		$
		H_{\min}\left(  B|X\right)  _{\rho}\equiv\inf_{x\in\mathcal{X}}H_{\min}\left(
		B\right)  _{\rho_{x}}$~\cite{R05}.
		\begin{align}
		%
		\Pi_{m_1,m_2}^{\prime\prime} &  \equiv 
		\Pi_{\bar{\rho},\delta}^{n} \cdot \PIonetwo = 
		\Pi_{\bar{\rho},\delta}^{n} \PIonetwo \Pi_{\bar{\rho},\delta}^{n}, \label{eq:proj-sandwitch-hmin} 
		\end{align}
		which allows us to bound the sum rate by an entropic quantity as in \eqref{eqn:sum-rate-bound}.
		Because the typical projectors $\PIone$ and $\PItwo$ are missing from
		the measurement, we cannot obtain von Neumann entropy bounds on the
		individual rates, but obtain bounds based on min-entropies \cite{FHSSW11}. 
}{}

	\subsection{Rate-splitting}
	\label{sec:mac-rate-splitting}

		Rate-splitting is another approach for achieving the classical multiple access channel rate region
		\cite{GRUW01}, which generalizes readily to the quantum case
		using the successive decoding approach in \cite{winter2001capacity}.
		
		\begin{lemma}
			For a given $p=p_{X_1},p_{X_2}$, any rate pair 
			$(R_1,R_2)$ that lies in between the two corner points 
			of the MAC rate region $\alpha_p$ and $\beta_p$
			can be achieved if Sender 2 splits her message $m_2$ into 
			two parts $m_{2u}$ and $m_{2v}$ and encodes them with a \emph{split codebook} 
			$\left(\{u^n(m_{2u})\}_{m_{2u}},\{v^n(m_{2v})\}_{m_{2v}},f\right)$.
			The receiver decodes the messages in the order $m_{2u} \to m_1 \to m_{2v}$
			using successive decoding.
		\end{lemma}

		The rate-split codebook consists of two random codebooks
		generated from $p_U$ and $p_V$ and a mixing function such that $f(U,V)=X_2$.
		For a fixed rate pair $(R_1, R_2)$, the construction of a split codebook
		achieving this rate pair depends on the properties of the channel for which we are coding.

\section{Quantum interference channels}
\label{sec:QIC}

	In this section 
	we calculate achievable rate regions for the quantum interference channel
	based on three decoding strategies: 	successive decoding,  simultaneous decoding
	and rate-splitting. 
	We also show the quantum Han-Kobayashi inner bound, 
	which relies on Conjecture~\ref{conj:sim-dec} for its proof.

	\subsection{Rates achievable by successive decoding}
	\label{sec:rate-succ-decoding}

		In this section, we require the receivers to decode the messages of both senders.
		Let the decoding ordering of Receiver 1 be represented by a permutation $\pi_1$:
		$\pi_1=(1,2)$ when decoding in the order $m_1 \to m_2$, 
		and $\pi_1=(2,1)$ for the alternate decoding
		order.
		%
		We similarly let $\pi_2=(1,2)$ and $\pi_2=(2,1)$ denote the two decode orderings
		for Receiver~2.
		%
		If we use a successive decoding strategy at both receivers,
		and calculate the best possible rates that are compatible 
		with both receivers' ability to decode, we obtain an achievable rate region.
		Consider, for example, the decoding strategy $\pi_1 = (2,1),\pi_2 = (2,1)$,
		which corresponds to both receivers decoding in the order $m_2 \to m_1$.
		In this case, we know that the code is decodable for Receiver~1 
		provided $R_1 < I(X_1; B_1|X_2)$ and $R_2 < I(X_2; B_1)$. 
		Receiver~2 will be able to decode provided $R_2 < I(X_2; B_2)$
		(we do not  require Receiver~2 to decode $m_1$ after he has decoded $m_2$).
		   
		%
		Thus, the rate pair $R_1 <I(X_1; B_1|X_2)$, 
		$R_2 <  \min\{I(X_2; B_1), I(X_2; B_2)\}$ is achievable for the interference 
		channel. Similarly, for all possible pairs of permutations $\pi_1, \pi_2$, we obtain
		an achievable rate pair for the interference channel.

		For interference channels with ``very strong'' interference \cite{carleial1975case},
		such that for all input distributions $p_{X_1}$ and $p_{X_2}$,
		\begin{align}	
			I\left(  X_{1};B_{1}|X_{2}\right)    &  \leq I\left(  X_{1}%
			;B_{2}\right) ,\label{eq:VSI-1}\\
			I\left(  X_{2};B_{2}|X_{1}\right)  &  \leq I\left(  X_{2}%
			;B_{1}\right),  \label{eq:VSI-2}%
		\end{align}
		the rates achieved by the successive decoding strategy $\pi_1 = (2,1),\pi_2 = (1,2)$
		are optimal.

		



		\begin{theorem}[Channels with very strong interference]
		\label{thm:carleial}
		The channel's capacity region is the union of 
		all rates $R_{1}$ and $R_{2}$ satisfying the inequalities:
		\begin{align*}
		R_{1}  &  \leq I\left(  X_{1};B_{1}|X_{2}Q\right)  _{\theta},\\
		R_{2}  &  \leq I\left(  X_{2};B_{2}|X_{1}Q\right)  _{\theta},
		\end{align*}
		with union taken over input distributions $p_Q$, $p_{X_{1}|Q}$ and $p_{X_{2}|Q}$.
		\end{theorem}

		%

		The  matching outer bound  follows from the converse part of Theorem~\ref{thm:cqmac-capacity},
		since the individual rates are optimal in the two MAC sub-channels \cite{carleial1975case}.
		Indeed, we can pursue the connection between the IC and the MAC sub-channels
		further to obtain a simple outer bound for the capacity of general quantum interference channels
		analogous to the classical result by Sato \cite{Sato77}.

		\begin{theorem}
		\label{thm:sato-weaker} Consider the Sato region defined as follows:
		\begin{equation}
		\mathcal{R}_{\text{Sato}}(\mathcal{N})\triangleq 
		\bigcup_{p \in \mcal{P}_{Sato}
		}\{(R_{1},R_{2})\}, \label{region:Gsato}%
		\end{equation}
		where $R_{1}$ and $R_{2}$ are rates satisfying the inequalities:
		\begin{align}
		R_{1}  &  \leq I(X_{1};B_{1}|X_{2}Q)_{\theta},\\
		R_{2}  &  \leq I(X_{2};B_{2}|X_{1}Q)_{\theta},\label{Gsato}\\
		R_{1}+R_{2}  &  \leq I(X_{1}X_{2};B_{1}B_{2}|Q)_{\theta},
		\end{align}
		where the union is taken over over all input distributions
		of the form $p_Q(q)\ p_{X_1|Q}(x_{1}|q)\ p_{X_2|Q}(x_{2}|q)$
		and the resulting average input-output state $\theta$.
		Then the region $\mathcal{R}_{\text{Sato}}$ is an outer bound on the
		capacity region of the general quantum interference channel.
		\end{theorem}
	
		This proof follows from the observation 
		that any code for the quantum interference channel also gives codes for three
		quantum multiple access channel subproblems: one for Receiver~1, another
		for Receiver~2, and a third for the two receivers considered together.
		Thus, using the outer bound on the quantum multiple access channel rates 
		from Theorem~\ref{thm:cqmac-capacity} we obtain the outer bound in 
		Theorem \ref{thm:sato-weaker}.

	\subsection{Rates achievable by two-sender simultaneous decoding}
	\label{sec:strong-int}

		The 
		simultaneous decoder from 
		Theorem~\ref{thm:sim-dec-two-sender} allows us to calculate the 
		capacity region for quantum interference channels with ``strong'' interference \cite{sato1981capacity,costa1987capacity},
		for which the following condition holds:
		\begin{align}
			I\left(  X_{1};B_{1}|X_{2}\right)    &  \leq  I\left(  X_{1}	;B_{2}|{X_{2}}\right) ,\label{eq:SI-1}\\
			I\left(  X_{2};B_{2}|X_{1}\right)  &  \leq    I\left(  X_{2}	;B_{1}|{X_{1}}\right),  \label{eq:SI-2}%
		\end{align}
		for all input distributions $p_{X_1}$ and $p_{X_2}$.
						
		\begin{theorem}[Channels with strong interference]
		\label{thm:strong-int}
		The channel's capacity region is the union of 
		all rates $R_{1}$ and $R_{2}$ satisfying the inequalities:
		\begin{align*}
		R_{1}  &  \leq I\left(  X_{1};B_{1}|X_{2}Q\right)  _{\theta},\\
		R_{2}  &  \leq I\left(  X_{2};B_{2}|X_{1}Q\right)_{\theta}, \\
		R_1 + R_{2}  &  \leq \min\{ I\left(X_{1}X_{2};B_{1}|Q\right), I\left(X_{1}X_{2};B_{2}|Q\right)_{\theta} \},
		\end{align*}
		where the union is over input distributions $p_{X_{1}|Q}\ p_{X_{2}|Q}\ p_Q$.
		\end{theorem}
		
		This rate region describes the intersection of the MAC rate regions
		for the two receivers and corresponds to the condition that we require each 
		receiver to decode both $m_1$ and $m_2$.

	\subsection{The quantum Han-Kobayashi rate region}
	\label{sec:HK}

		For general interference channels the Han-Kobayashi coding strategy 
		gives the best known achievable rate region \cite{HK81} and involves
		\emph{partial} decoding of the interfering signal.
		Instead of using a standard codebook $\{x^n_1(m_1)\}_{m_1\in \mathcal{M}_1}$
		at a rate $R_1 \equiv \frac{1}{n}\log|\mathcal{M}_1|$ to encode her message $m_1$,
		Sender~1 splits  her message into two parts: a \emph{personal} message $m_{1p}$ 
		encoded using a random  codebook $\{u^n_1(m_{1p})\}_{m_{1p}\in \mathcal{M}_{1p}}$
		and a common message $m_{1c}$ encoded into 
		$\{w^n_1(m_{1c})\}_{m_{1c}\in \mathcal{M}_{1c}}$. 
		In terms of rates, this means that the sum rate $R_{1p}+R_{1c}$ should 
		be equal to the original rate $R_1$. 
		So long as Receiver~1 can decode both parts $m_{1p}$ and $m_{1c}$,
		he can reconstruct the original message $m_1$.
		Receiver~2 can decode Sender~1's common message $m_{1c}$ and 
		improve his communication rate to $R_2=I(X_2;B_2|W_1)$ by using this
		side information.
		To return the favor, Sender~2 also splits her message into $m_{2p}$ and $m_{2c}$.
		The overall codebook is generated from the class of Han-Kobayashi probability distributions, 
		$\mathcal{P}_{HK}$, which factorize as
		$p(q)p(u_1|q)p(w_1|q)p(x_1|u_1,w_1)
		 p(u_2|q)p(w_2|q)p(x_2|u_2,w_2)$,
		where $p(x_1|u_1,w_1)$ and $p(x_2|u_2,w_2)$ are degenerate
		probability distributions that correspond to deterministic functions $f_1$ and $f_2$,
		$f_i\colon\mathcal{U}_i\times\mathcal{W}_i \to \mathcal{X}_i$, which are used to 
		combine symbols of $U$ and $W$ to produce a symbol $X$ suitable
		as input to the channel.

		\begin{theorem}
		\label{thm:quantum-HK-region}
		The quantum Han-Kobayashi rate region:
		        \begin{equation}
		        		\nonumber
		        		\HKR 
					\equiv 
						\bigcup_{ p \in \mathcal{P}_{HK} } 
						\{ (R_1,R_2) \in \bbR^2 | \text{ \emph{Eqns. (HK1) - (HK9) }} \} 
		        \end{equation}
		        {\small
		        \begin{align}
		            R_1 		&\leq		I(U_1W_1;B_1|W_2Q)   							\tag{HK1} \\
		            R_1 		&\leq		I(U_1;B_1|W_1W_2Q) + I(W_1;B_2|U_2W_2Q)	            \tag{HK2} \\
		            R_2 		&\leq		 \qquad \qquad  \qquad \qquad \ \ \ \ I(U_2W_2;B_2|W_1Q)    							\tag{HK3} \\
		            R_2 		&\leq		 I(W_2;B_1|U_1W_1Q) + I(U_2;B_2|W_1W_2Q) 		\tag{HK4} \\
		            R_1 + R_2	&\leq		I(U_1W_1W_2;B_1|Q) +  I(U_2;B_2|W_1W_2Q)		\tag{HK5}\\
		            R_1 + R_2	&\leq		  I(U_1;B_1|W_2W_1Q) +I(U_2W_2W_1;B_2|Q) 	\tag{HK6} \\
		            R_1 + R_2	&\leq		I(U_1W_2;B_1|W_1Q)   + I(U_2W_1;B_2|W_2Q) 		\tag{HK7} \\
		            2R_1 + R_2	&\leq		I(U_1;B_1|W_1W_2Q) +  I(U_2W_1;B_2|W_2Q)
		            					\nonumber \\ 			& 		\quad 
								\!\!+ \! I(U_1W_1W_2;B_1|Q)  	 \tag{HK8} \\
		            R_1 + 2R_2	&\leq		I(U_1W_2;B_1|W_1Q) + I(U_2;B_2|W_2W_1Q) 
		            					\nonumber \\  	&		\qquad \qquad	\qquad \qquad \quad \ \
								+  I(U_2W_2W_1;B_2|Q)		\tag{HK9} 
		        \end{align}}where the information theoretic quantities are taken with respect 
			to a state of the form:
			\vspace{-2mm}
			\begin{align}
			& 
				\sum_{q,u_{1},u_{2},w_{1},w_{2}	}
				\hspace*{-5mm}
				p_{Q}\!\left(  q\right)  
				p_{U_{1}}\!\left(  u_{1}|q\right)  
				p_{U_{2}}\!\left(  u_{2}|q\right)  
				p_{W_{1}}\!\left(  w_{1}|q\right)  
				p_{W_{2}}\!\left(  w_{2}|q\right) \qquad \qquad \nonumber \\[-1mm]
			&  \qquad \quad
				\left\vert q\right\rangle \! \left\langle q\right\vert ^{Q}
				\otimes
				\left\vert u_{1}\right\rangle \! \left\langle u_{1}\right\vert ^{U_{1}}
				\otimes
				\left\vert u_{2}\right\rangle \! \left\langle u_{2}\right\vert ^{U_{2}} 
				\otimes	
					\label{eq:ipsi-EA-interference-state}\\
			&\quad \qquad  \quad
				\left\vert w_{1}\right\rangle \! \left\langle w_{1}\right\vert ^{W_{1}}
				\otimes 
				\left\vert w_{2}\right\rangle \! \left\langle w_{2}\right\vert ^{W_{2}}
				\otimes
				\rho_{f_{1}\left(  u_{1},w_{1}\right)  ,f_{2}\left(  u_{2},w_{2}\right)  }^{B_{1}B_{2}}, \nonumber 
			\end{align}
			is an achievable rate region provided Conjecture~\ref{conj:sim-dec} holds.			



		\end{theorem}
		
		The proof is in the same spirit as the original result of Han and Kobayashi \cite{HK81}.
		%
		Our result is conditional on 
		Conjecture~\ref{conj:sim-dec} 
		for the construction of the decoding POVMs:
		$\left\{\Lambda_{m_{1p},m_{1c},m_{2c}} \right\}$
		for Receiver~1, and $\left\{  \Gamma_{m_{1c},m_{2c},m_{2p}}  \right\} $ 
		for Receiver~2.
		Refer to \cite{FHSSW11} for the proof.
		%

		\begin{figure}
		[ptb]
		\begin{center}
		\hspace*{-0.1cm}\includegraphics[
		natheight=1.932900in,
		natwidth=4.540300in,
		width=3.5in
		]%
		{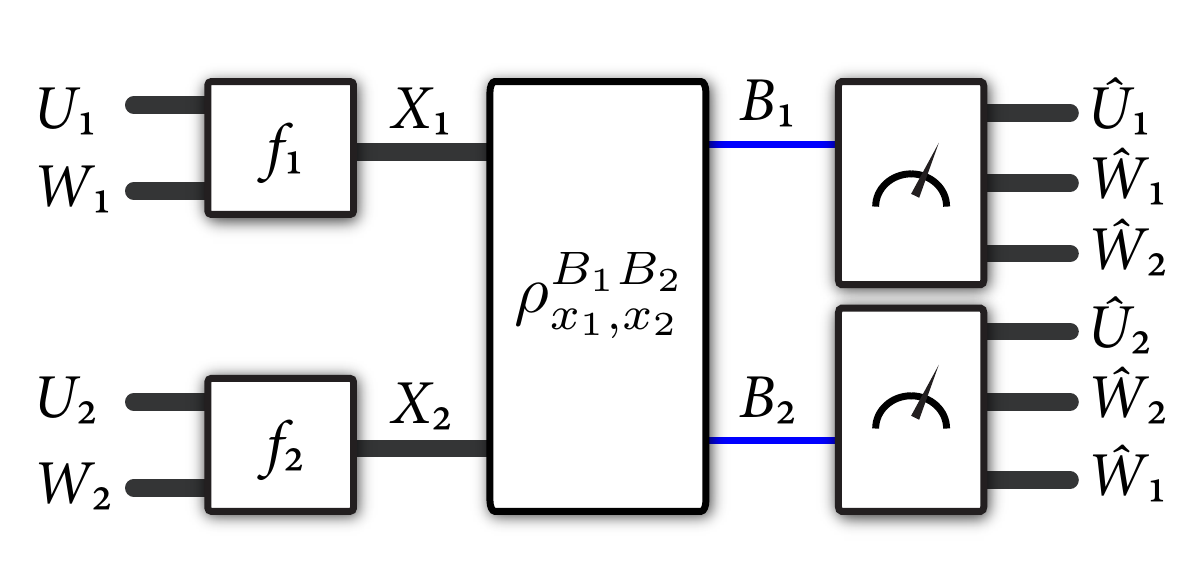}%
		\caption{The Han-Kobayashi coding strategy. Sender~1 selects codewords
		according to a \textquotedblleft personal\textquotedblright\ random variable
		$U_{1}$ and a \textquotedblleft common\textquotedblright\ random variable
		$W_{1}$. She then acts on $U_{1}$ and $W_{1}$ with some deterministic function $f_{1}$ that
		outputs a variable $X_{1}$ which serves as a classical input to the
		interference channel. Sender~2 uses a similar encoding. Receiver~1 performs a
		measurement to decode both variables of Sender~1 and the common random
		variable $W_{2}$ of Sender~2. Receiver~2 acts similarly. 
		}%
		\label{fig:han-kob-code}%
		\end{center}
		\end{figure}

\ifthenelse{\boolean{WITHHMINS}}{

}{}

	\subsection{Using rate-splitting for the IC}
	\label{sec:rate-succ-decoding-rs}

		We can use rate-splitting to improve the successive decoding region described in Section \ref{sec:rate-succ-decoding}. 
		Inspired by the Han-Kobayashi strategy we make the senders split their messages into two parts:
		$m_1 \to m_{1p},m_{1c}$ and $m_2 \to m_{2p},m_{2c}$.
		Such a split induces two three-user multiple access  channels.
		Receiver~1 decodes the messages $m_{1p}, m_{1c}$ and $m_{2c}$ using successive decoding,
		and there are six different decode orderings he can use.
		We can naturally use all  $6 \times 6$ pairs of decoding orders to obtain a set of achievable 
		rate pairs. 

		\begin{proposition}
			Consider the rate point $P$ associated with the decode ordering $\pi_1$ for Receiver~1 and
			$\pi_2$ for Receiver~2:
			\be
				P=\left(R^{(1)}_{1p} + \min\{R^{(1)}_{1c},R^{(2)}_{1c}\}, \  \min\{R^{(1)}_{2c},R^{(2)}_{2c}\} + R^{(2)}_{2p} \right), 
				\nonumber
			\ee
			where the rates constraints for Receiver $j$ satisfy
			%
		        \begin{align}        	 
				R^{(j)}_{\pi_j(1)}		&\leq I(X_{\pi_j(1)};B_j), \\
				R^{(j)}_{\pi_j(2)}		&\leq I(X_{\pi_j(2)};B_j|X_{\pi_j({1})}), \\
				R^{(j)}_{\pi_j(3)}		&\leq I(X_{\pi_j(3)};B_j|X_{\pi_j({1})}X_{\pi_j({2})}).
		        	\end{align}        
			The rate pair $P$ is achievable for the quantum interference channel,
			for all permutations $\pi_1$ of the set of indices  $(1p,1c,2c)$ 
			and for all permutations $\pi_2$ of the set  $(2p,2c,1c)$.
		\end{proposition}

			\begin{figure}[ptb]
			\begin{center}
			\hspace*{-0.4cm}\includegraphics[width=3.8in]{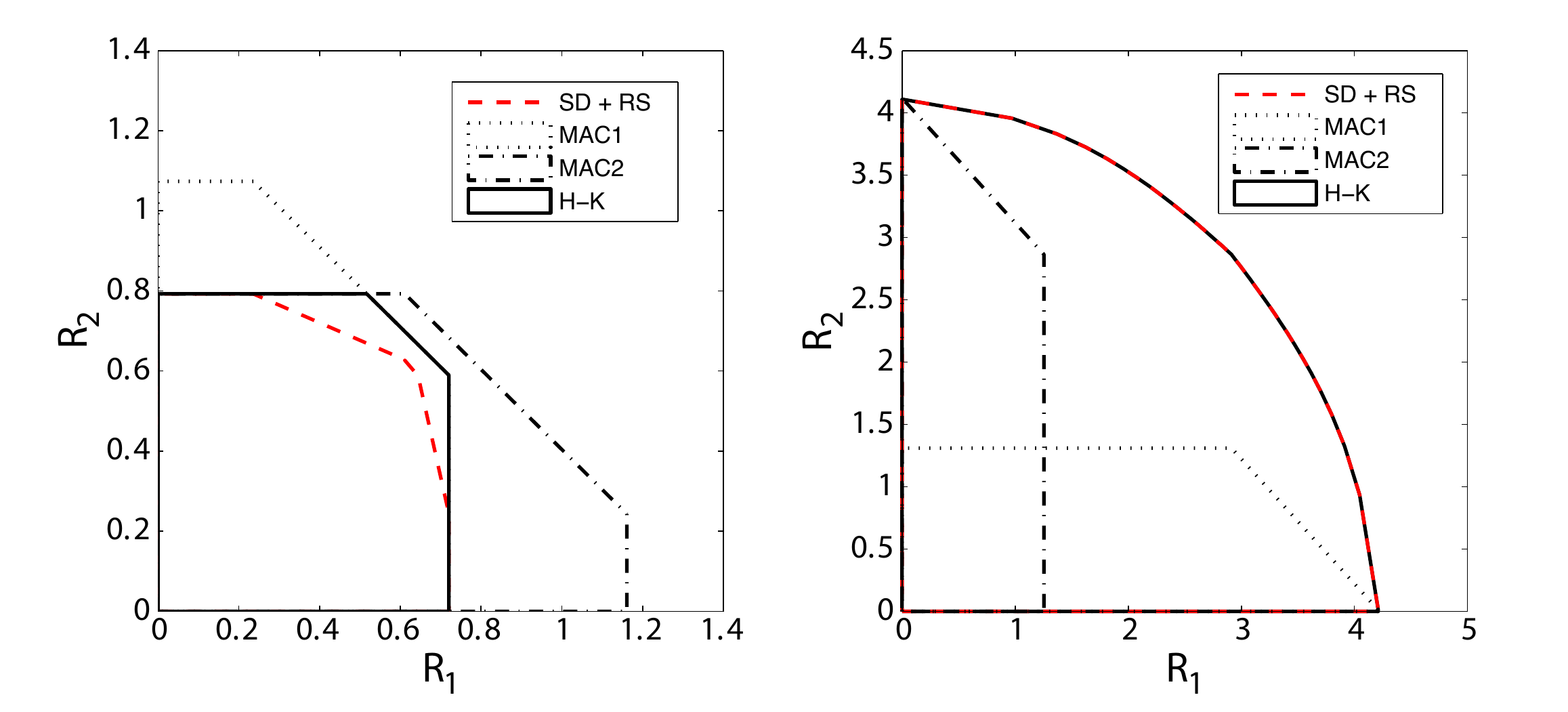}%
			\caption{
			These two figures plot rate pairs that are achievable with successive decoding and rate-splitting (SD+RS). 
			The figures compare these rates with those achievable by the Han-Kobayashi (HK) coding strategy, while also plotting
			the regions corresponding to the two induced multiple access channels to each
			receiver (MAC1 and MAC2). 
			The LHS figure demonstrates that 
			%
			SD+RS does not in general achieve the full capacity region 
			for channels with strong interference.
			For this case we can use the two-sender simultaneous decoder from 
			Theorem~\ref{thm:sim-dec-two-sender}.
			The RHS figure demonstrates that, for some channels with weak interference, SD+RS is
			virtually indistinguishable from HK.
		%
		%
				%
				%
				%
				%
				%
				%
			}%
			\label{fig:jt-succ-decoder-rate-splitting}%
			\end{center}
			\end{figure}

		The rate region described by the convex hull of the points $P$
		is generally smaller than the Han-Kobayashi region as illustrated in Figure~\ref{fig:jt-succ-decoder-rate-splitting}.
		An interesting open problem is whether we can achieve all rates of the Han-Kobayashi region
		  by splitting each sender's message
		into more than two parts and using only rate-splitting \cite{GRUW01} and successive
		decoding.
		There exists an attempt to answer this question for the classical interference channel~\cite{sasoglu2008successive}.
		%
		The argument in that paper is based on a careful analysis of the geometrical 
		structure of the Chong-Motani-Garg region,
		which is known to be equivalent to the Han-Kobayashi region when
		considering all possible input distributions~\cite{CMGE08}.
		An implicit assumption is made that 
		the change of the code distribution dictated by applying
		the rate-splitting technique at the convenience of one receiver does not
		affect the other receiver's decoding ability.  
		Unfortunately, this assumption does not hold in general,
		which can be seen from the following argument.

		Consider a code for an interference channel where the message $m_1\in\{1,\ldots,2^{nR}\}$ 
        is to be decoded by both receivers. 
        Suppose we have $R=I(X_1;Y_{2})$ and $R\leq	I(X_1;Y_{1})$ for some input distribution $p_{X_1}$.
		If we generate a standard random codebook of size $2^{nR}$,  then both receivers will be able to decode the message
        encoded in $X_1$. 
		However, 
		we might want to use a split codebook generated according to 
		distributions $p_{U}$ and $p_{V}$, and the mixing function $f(U,V)=X_1$.
		If we generate the split codebook \emph{for} Receiver~2 then we should pick 
		the rate $R_{U}=I(U;Y_{2})$ 
		so that Receiver~2 will be able to decode $U$ 
		with small error probability. 
		We should however keep	in mind that we are coding for an interference channel and we also want
		Receiver$~1$ to decode $X_1$. 
		The problem is that it is possible that $R_{U} >I(U;Y_{1})$, 
		in which case Receiver$~1$ cannot decode $U$ 
		and thus cannot decode the message by successive decoding. In this case, the code obtained by splitting
		according to the second receiver's prescription is not a good code for the
		interference channel.

\section{Discussion}
\label{sec:discussion}
	

	There are several open questions regarding this work.
	First, we would of course like to prove Conjecture~\ref{conj:sim-dec}
	holds  because it would be a powerful building block
	for multi-user quantum Shannon theory. 
	%
	Also, we would like to study the channel's quantum, entanglement-assisted,
	 and hybrid classical-quantum capacities.
	 Finally, it could be that three-sender quantum simultaneous decoding is not
	 necessary for achieving the Han-Kobayashi region. 
	If the classical Han-Kobayashi rate region for the discrete memoryless interference channel 
	can be achieved using rate-splitting and successive decoding, then
	this would be another way to prove Theorem~\ref{thm:quantum-HK-region}
	without appealing to Conjecture~\ref{conj:sim-dec}.

	We acknowledge discussions with Fr\'{e}d\'{e}ric Dupuis,  
	Eren \c{S}a\c{s}o\u{g}lu and Mai Vu. 
	P.~Hayden acknowledges support from the Canada Research Chairs program, the Perimeter
	Institute, CIFAR, FQRNT's INTRIQ, MITACS, NSERC, ONR through grant
	N000140811249, and QuantumWorks. M.~M.~Wilde acknowledges support from the
	MDEIE (Qu\'{e}bec) PSR-SIIRI international collaboration grant.
	I.~Savov acknowledges  support from FQRNT and NSERC.

\ifthenelse{\boolean{WITHAPDX}}{

\appendix

\subsection{Typical Sequences and Typical Subspaces} 		\label{sec:typ-review}

	
	We present here a number of properties of typical sequences and their quantum 
	analogue: typical subspaces.
	
	\medskip 
	
	\noindent
	{\bf Classical typicality\ \ }
	Denote by $x^n$ a sequence $x_1x_2\dots x_n$, where each
	$x_i, i\in[n]$ belongs to the finite \emph{alphabet} $\cX$. 
	Denote by $|\cX|$ the cardinality of $\cX$. 
	%
	To avoid confusion, we use $i\in[n]$ to denote the index of a symbol $x$
	in the sequence $x^n$ and $a \in [1,2,\ldots,|\mcal{X}|]$ 
	to denote the different symbols in the alphabet $\mcal{X}$.

\def\cT{{\mathcal A}}

	Consider the random variable $X$ with probability distribution $p_{X}(x)$ defined on a finite set $\cX$.
	Let $H(X) \equiv H(p_X) \equiv - \sum_x p_X(x) \log p_X(x)$ be the Shannon entropy of $p_X$. 
	Define the probability distribution $p_{X^n}(x^n)$ on $\cX^n$ to be the
	$n$-fold product of $p_{X}$. 
	The sequence $x^n$ is drawn from $p_{X^n}$ if
	and only if each letter $x_i$ is drawn independently from $p_X$.
	For any $\delta >0$,	define the set of entropy $\delta$-typical sequences of length 
	$n$ as: 
	\be
		\cT^{n}_{p_X,\delta} 
		\! \equiv \!
		\left\{ \!
			x^n \in \mcal{X}^n  \colon \! \left|  -\frac{\log p_{X^n}(x^n)}{n}  - H(X) \right| \! \leq \delta 
		\right\}.%
	\ee

	Typical sequences enjoy many useful properties \cite{CT91}.
	For any $\epsilon,\delta>0$, and sufficiently large $n$, we have
	\begin{eqnarray}
		& 
		 \!\!\!\!\!\!\!\!\!\!\!\!\!\!\!\!\!
		\displaystyle\sum_{x^n\in \cT_{p_X,\delta}^{n}} \!\!\!\! p_{X^n}\!\!\left( x^n\right)  
		& \geq  1-\epsilon,   \label{cc1}  \\[1.6mm]
		2^{-n[ H(X)+\delta ]}  \leq &
		p_{X^n}(x^n) & 
		\leq   2^{-n[H(X)-\delta]}   \nonumber \\[-1mm]
		& & \hspace*{-10mm}  \forall x^n \in \cT_{p_X,\delta}^{n},  \label{cc2} \\[2.2mm]
		[1 - \epsilon] 2^{n[ H(X)-\delta]} \leq &  
		|\cT_{p_X,\delta}^{n}| 
		& \leq 2^{n[ H(X)+\delta]}. \label{cc3} 
	\end{eqnarray}

	\bigskip

	\noindent
	{\bf Quantum typicality\ \ }
	The above concepts generalize to the quantum setting by virtue of the spectral
	theorem. Let $\cH^B$ be a $d_B$ dimensional Hilbert space and let 
	$\rho^B  \in \mcal{D}(\cH^B)$ be the density matrix associated with a quantum state.
	The spectral decomposition of $\rho^B$ is denoted
	$\rho^B=U\Lambda U^\dag$ where $\Lambda$ is a diagonal
	matrix of positive real eigenvalues that sum to one.
	We identify the eigenvalues of $\rho^B$ with the probability 
	distribution $p_Y(y)=\Lambda_{yy}$ and write the 
	spectral decomposition as:
	\be
		\rho^B = \sum_{y=1}^{d_B} p_Y(y) \ket{e_{\rho;y}}\bra{e_{\rho;y}}^B
		\label{eq:spectral-decomp-rho}
	\ee
	where $\ket{e_{\rho;y}}$ is the eigenvector of $\rho^B$ corresponding to eigenvalue $p_Y(y)$.
	The von Neumann entropy of the density matrix $\rho^B$ is
	\be
		H(B)_\rho=-\Tr\{\rho^B\log\rho^B\}=H(p_Y).
	\ee

	Define the set of  $\delta$-typical eigenvalues according to the eigenvalue distribution $p_Y$
	\be
		\cT^{n}_{p_Y,\delta} 
		\! \equiv \!
		\left\{ \!
			y^n \in \mcal{Y}^n  \colon \! \left|  -\frac{\log p_{Y^n}(y^n)}{n}  - H(Y) \right| \! \leq \delta 
		\right\}.%
	\ee
	For a given string $y^n = y_1y_2\ldots y_i\ldots y_n$ we define the 
	corresponding eigenvector as 
	\be
		\ket{e_{\rho;y^n}} = \ket{e_{\rho;y_1}} \otimes \ket{e_{\rho;y_2}} \otimes \cdots \otimes \ket{e_{\rho;y_n}},
	\ee
	where for each symbol where $y_i=b \in \{1,2,\ldots,d_B\}$ we select the b$^\textrm{th}$ eigenvector 
	$\ket{e_{\rho;b}}$.
	
	The typical subspace associated with the density matrix $\rho^B$
	is defined as
	\be
	{A}^n_{\rho, \delta}
		= \textrm{span} \!  \{  \ket{e_{\rho;y^n}} \colon y^n \in \cT_{p_Y,\delta}^n  \}.
	\ee
	The typical projector is defined as 
	\be
		\Pi^{n}_{\rho^B, \delta}
		= \sum_{y^n\in\cT_{p,\delta}^n}	    \ket{e_{\rho;y^n}}\! \bra{e_{\rho;y^n}}.
	\ee
	Note that the typical projector is linked twofold to the spectral decomposition
	of \eqref{eq:spectral-decomp-rho}: the sequences $y^n$ are selected 
	according to $p_Y$ and the set of typical vectors are build from tensor
	products of orthogonal eigenvectors $\ket{e_{\rho;y}}$.

	Properties analogous to (\ref{cc1}) -- (\ref{cc3}) hold.
	For any $\epsilon,\delta>0$, and all sufficiently large $n$ we have

	\vspace{-4mm}
	{\footnotesize
	\begin{eqnarray}
		\label{eqn:TypP-prop-one}
		&\!\!\!\!\!\!\!\Tr\{\rho^{\otimes n} \Pi^{n}_{\rho, \delta}\}  &\geq  1-\epsilon  \\
		\label{eqn:TypP-prop-two}
		2^{-n[ H(B)_\rho+\delta]}\Pi^n_{\rho, \delta} 
		\leq  
		&\!\!\Pi^n_{\rho, \delta} \rho^{\otimes n} \Pi^{n}_{\rho, \delta}
		\!\!&
		\leq 
		2^{-n[ H(B)_\rho-\delta ]}\Pi^{n}_{\rho, \delta}, \\
	\label{eqn:TypP-prop-three}
		[1 - \epsilon] 2^{n[ H(B)_\rho-\delta]}
		\leq  
		&\Tr\{\Pi^{n}_{\rho, \delta}\} &
		\leq 
		2^{n[ H(B)_\rho+\delta]}.
	\end{eqnarray}%
	\noindent
	}
	The interpretation of \eqref{eqn:TypP-prop-two} is that the eigenvalues 
	of the  state $\rho^{\otimes n}$ are bounded between
	$2^{-n[ H(B)_\rho-\delta ]}$ and $2^{-n[ H(B)_\rho+\delta ]}$
	on the typical subspace ${A}^n_{\rho, \delta}$.
	
	\bigskip 
	
	\noindent
	{\bf Signal states\ \ }
	Consider now a set of quantum states $\{\rho_{x_a}\}$,
	$x_a \in \mcal{X}$.
	We perform the spectral decomposition of each $\rho_{x_a}$
	to obtain 
	\be
		\rho^B_{x_a}= \sum_{y=1}^{d_B} p_{Y|X}(y|x_a) \ket{e_{\rho_{x_a};y}}\bra{e_{\rho_{x_a};y}}^B,
		\label{eq:cond-spectral-decomp-rho}
	\ee
	where $p_{Y|X}(y|x_a)$  is the $y^\textrm{th}$ eigenvalue of  $\rho^B_{x_a}$ and 
	$\ket{e_{\rho_{x_a};y}}$ is the the corresponding eigenvector.
	
	We can think of $\{\rho_{x_a}\}$ as a classical-quantum (\emph{c-q}) channel
	where the input is some $x_a \in \mcal{X}$ and the output is the 
	corresponding quantum state $\rho_{x_a}$.
	If the channel is memoryless, then for  each input sequence 
	$x^n=x_1x_2\cdots x_n$ we have the corresponding tensor product output state: 
	\be
		\rho_{x^n} = \rho_{x_1} \otimes \rho_{x_2} \otimes \cdots \otimes \rho_{x_n}.
	\ee


	\medskip
	\noindent
	{\bf Conditionally typical projector \ \ }
	Consider the ensemble $\left\{  p_{X}\!\left(  x_a\right)  ,\rho_{x_a}\right\}$.
	The choice of distributions induces the following classical-quantum state:
	\be
		\rho^{XB} =
	       	\sum_{x_a} 
			p_{X}\!\left(x_{a}\right)
			\ketbra{x_a}{x_a}^{X} \!\!
			\otimes \!
			\rho^{B}_{x_a}.
	\ee

	We can now define the conditional entropy of this state as
	\be
		H (B|X)_\rho
		\equiv
		 \sum_{x_a \in \mcal{X}} 
			p_{X}(x_a) 
 			H (\rho_{x_a}),
	\ee	
	or equivalently, expressed in terms of the eigenvalues of the signal states,
	the conditional entropy becomes
	\be
		H (B|X)_\rho
		\equiv
		H(Y|X)
		\equiv
		\sum_{x_a}p_{X}(x_a)H(Y|x_a),
	\ee	
	where $H(Y|x_a) = - \sum_y p_{Y|X}(y|x_a) \log p_{Y|X}(y|x_a)$
	is the entropy of the eigenvalue distribution shown in \eqref{eq:cond-spectral-decomp-rho}.

	We define the $x^n$-conditionally typical projector
	as follows:
	\be
		\Pi^{n}_{\rho^B_{x^n}, \delta}
		=  \sum_{y^n \in \cT^{n}_{\rho^{B^n}_{x^n},\delta} }
			\ket{e_{\rho_{x^n};y^n}}\!\bra{e_{\rho_{x^n};y^n}},
		\label{eqn:cond-typ-projector-def}
	\ee
	where the set of conditionally typical eigenvalues $\cT^{n}_{\rho^{B^n}_{x^n},\delta}$
	consists of 
	all sequences $y^n$ which satisfy: 
	\be
		\cT^{n}_{\rho^{B^n}_{x^n},\delta} 
		\! \equiv \!
		\left\{ \!
			y^n 
			\colon \! \left|  -\frac{\log p_{Y^n|X^n}(y^n|x^n)}{n}  - H(Y|X) \right| \! \leq \delta 
		\right\},%
	\ee	
	with $p_{Y^n|X^n}(y^n|x^n) = \prod_{i=1}^n p_{Y|X}(y_i|x_i)$.
	
	%
	The states $\ket{e_{\rho_{x^n};y^n}}$ are built from tensor products of eigenvectors
	for the individual signal states:
	\be
		\ket{e_{\rho_{x^n};y^n}} = \ket{e_{\rho_{x_1};y_1}} \otimes \ket{e_{\rho_{x_2};y_2}} \otimes \cdots \otimes \ket{e_{\rho_{x_n};y_n}},
	\ee
	where the string $y^n = y_1y_2\ldots y_i\ldots y_n$ varies over different choices of bases for $\mcal{H}^B$.
	For each symbol  $y_i=b \in \{1,2,\ldots,d_B\}$ we select $\ket{e_{\rho_{x_a};b}}$:
	the b$^\textrm{th}$ eigenvector from the eigenbasis of $\rho_{x_a}$ corresponding to the letter $x_i = x_a \in \mathcal{X}$.
	%

	The following bound on the size of the conditionally typical projector
	applies:
	\be
		\Tr\{  \Pi^{n}_{\rho^B_{x^n}, \delta} \} \leq 2^{n[H(B|X)_\rho + \delta] }.
		\label{eqn:bound-on-size}
	\ee


	\bigskip
	\noindent
	{\bf MAC code \ \ }
	Consider now a quantum multiple access channel 
	$(\mathcal{X}_1 \times \mathcal{X}_2,  \rho_{x_1,x_2}^{B}, \mathcal{H}^{B} )$
	and two input distributions $p_{X_1}$ and $p_{X_2}$.
	Define the random codebooks $\{X_1^n(m_1)\}_{m_1 \in \cM_1}$ and $\{X_2^n(m_2)\}_{m_2 \in \cM_2}$
	generated from the product distributions $p_{X_1^n}$ and $p_{X_2^n}$ respectively. 
	The choice of distributions induces the following classical-quantum state $\rho^{X_{1}X_{2}B}$
	\be
	       	\sum_{x_a,x_b} 
			p_{X_{1}}\!\left(x_a\right)p_{X_{2}}\!\left( x_b\right)
			\ketbra{x_a}{x_a\!}^{X_1} \!\!
			\otimes \!
			\ketbra{x_b}{x_b}^{X_2} \!
			\otimes \!
			\rho^{B}_{x_ax_b}.
	\ee	        	       
	and the averaged output states:
	\begin{align}
		\bar{\rho}_{x_a} &  \equiv
			\sum_{x_b}p_{X_2}\!\left(  x_b\right)  \rho_{x_a,x_b},\label{eq:rho_xap} \\
		\bar{\rho}_{x_b} &  \equiv
			\sum_{x_a}p_{X_1}\!\left(  x_a\right)  \rho_{x_a,x_b},\label{eq:rho_yap} \\
		\bar{\rho} &  \equiv
			\sum_{x_a,x_b}p_{X_1}\!\left(  x_a\right)  p_{X_2}\!\left(  x_b\right) \rho_{x_a,x_b}.\label{eq:rhoap}
	\end{align}
	
	The conditional quantum entropy $H(B|X_1X_2)_\rho$ is:
	\be
		H (B|X_1X_2)_\rho
		= \hspace*{-5mm}
		 \sum_{x_a \in \mcal{X}_1, x_b \in \mcal{X}_2} 
		 \hspace*{-5mm}
			p_{X_1}(x_a) 
			p_{X_2}(x_b) 
			H (\rho_{x_a,x_b}),
	\ee
	and using the average states we define:
	\begin{align}
		H (B|X_1)_\rho
		&= 
		 \sum_{x_a \in \mcal{X}_1} 
			p_{X_1}(x_a) 
			H (\bar{\rho}_{x_a}),
	\\
		H (B|X_2)_\rho
		&=
		 \sum_{x_b \in \mcal{X}_2} 
			p_{X_2}(x_b) 
			H (\bar{\rho}_{x_b}),
	\\
		H (B)_\rho
		&=
		H(\bar{\rho}).
	\end{align}

	Similarly to equation \eqref{eqn:cond-typ-projector-def} and for each message pair $(m_1,m_2)$
	we define the conditionally typical projector for the encoded state $\rho^B_{x_1^n(m_1)x_2^n(m_2)}$ to be $\Pi^{n}_{\rho^B_{x_1^n(m_1)x_2^n(m_2)}, \delta}$.
 	From this point on, we will not indicate the messages $m_1$, $m_2$ explicitly, 
	because the codewords are constructed identically for each message.
	
	Analogous to \eqref{eqn:bound-on-size}, the following upper bound applies:
	\be
		\Tr\{  \Pi^{n}_{\rho^B_{x_1^nx_2^n}, \delta}  \} \leq 2^{n[H(B|X_1X_2)_\rho + \delta] },
	\ee
	and we can also bound from below the eigenvalues of the state 
	$\rho^B_{x_1^nx_2^n}$ as follows:
	\be
		2^{-n[H(B|X_1X_2)_\rho + \delta] } \Pi^{n}_{\rho^B_{x_1^nx_2^n}, \delta}  
		\leq 
		\Pi^{n}_{\rho^B_{x_1^nx_2^n}, \delta}  
		\rho^B_{x_1^nx_2^n}
		\Pi^{n}_{\rho^B_{x_1^nx_2^n}, \delta}.
	\ee
	

	We define conditionally typical projectors for each of the
	averaged states:
	\begin{align}
		\bar{\rho}_{x_1} &  \to
			\Pi^{n}_{\bar{\rho}^B_{x_1^n}, \delta},  \\
		\bar{\rho}_{x_2} &  \to
			\Pi^{n}_{\bar{\rho}^B_{x_2^n}, \delta}, \\
		\bar{\rho} &  \to
			\Pi^{n}_{\bar{\rho}^B, \delta}.
	\end{align}
	These projectors obey the standard eigenvalue upper bounds when
	acting on the states with respect to which they are defined:
	\begin{align}
			\Pi^{n}_{\bar{\rho}^B_{x_1^n}, \delta}
			\bar{\rho}_{x_1^n}
			\Pi^{n}_{\bar{\rho}^B_{x_1^n}, \delta}
		& \leq
			2^{-n[H(B|X_1)_\rho - \delta] }
			\Pi^{n}_{\bar{\rho}^B_{x_1^n}, \delta},  \\
			\Pi^{n}_{\bar{\rho}^B_{x_2^n}, \delta}	
			\bar{\rho}_{x_2^n}
			\Pi^{n}_{\bar{\rho}^B_{x_2^n}, \delta}
		& \leq
			2^{-n[H(B|X_2)_\rho - \delta] } 
			\Pi^{n}_{\bar{\rho}^B_{x_2^n}, \delta},  \\			
			\Pi^{n}_{\bar{\rho}^B, \delta} \ 
			\bar{\rho}^B \
			\Pi^{n}_{\bar{\rho}^B, \delta}
		& \leq 
			2^{-n[H(B)_\rho - \delta] }
			\Pi^{n}_{\bar{\rho}^B, \delta}.
	\end{align}	

	The encoded state $\rho^B_{X_1^nX_2^n}$ is well supported by
	all the typical projectors on average:
	\begin{align}
		\mathbb{E}_{X^n_1X^n_2} \left[
		\Tr\{
			\Pi^{n}_{\rho^B_{X_1^nX_2^n}, \delta} \
			\rho^B_{X_1^nX_2^n}
		\}
		\right]		
		& \geq 1 - \epsilon,		\\
		\mathbb{E}_{X^n_1X^n_2} \left[
		\Tr\{
			\Pi^{n}_{\bar{\rho}^B_{X_1^n}, \delta} \
			\rho^B_{X_1^nX_2^n}
		\}
		\right]
		& \geq 1 - \epsilon,		\\
		\mathbb{E}_{X^n_1X^n_2} \left[
		\Tr\{
			\Pi^{n}_{\bar{\rho}^B_{X_2^n}, \delta} \
			\rho^B_{X_1^nX_2^n}			
		\}
		\right]
		& \geq 1 - \epsilon, 	\\
		\mathbb{E}_{X^n_1X^n_2} \left[
		\Tr\{
			\Pi^{n}_{\bar{\rho}^B, \delta} \
			\rho^B_{X_1^nX_2^n}
		\}
		\right]		
		& \geq 1 - \epsilon.
	\end{align}
		
	

	\bigskip

	\noindent 
	Finally, we state this useful lemma: 
		
		\begin{lemma}[Gentle Operator Lemma for Ensembles \cite{itit1999winter,ON07}]\label{lem:gentle-operator}
		Given an ensemble $\left\{  p_{X}\left(  x\right)  ,\rho_{x}\right\}  $ with
		expected density operator $\rho\equiv\sum_{x}p_{X}\left(  x\right)  \rho_{x}$,
		suppose that the operator $\Lambda$ such that $0 \leq \Lambda\leq I$ succeeds
		with high probability on the state $\rho$:%
		\[
		\text{Tr}\left\{  \Lambda\rho\right\}  \geq1-\epsilon.
		\]
		Then the subnormalized state $\sqrt{\Lambda}\rho_{x}\sqrt{\Lambda}$ is close
		in expected trace distance to the original state $\rho_{x}$:%
		\[
		\mathbb{E}_{X}\left\{  \left\Vert \sqrt{\Lambda}\rho_{X}\sqrt{\Lambda}%
		-\rho_{X}\right\Vert _{1}\right\}  \leq2\sqrt{\epsilon}.
		\]

		\end{lemma}



\bigskip

}
{}

\bibliographystyle{IEEEtran}
\bibliography{interferenceChannel}

\begin{thebibliography}{10}
\providecommand{\url}[1]{#1}
\csname url@samestyle\endcsname
\providecommand{\newblock}{\relax}
\providecommand{\bibinfo}[2]{#2}
\providecommand{\BIBentrySTDinterwordspacing}{\spaceskip=0pt\relax}
\providecommand{\BIBentryALTinterwordstretchfactor}{4}
\providecommand{\BIBentryALTinterwordspacing}{\spaceskip=\fontdimen2\font plus
\BIBentryALTinterwordstretchfactor\fontdimen3\font minus
  \fontdimen4\font\relax}
\providecommand{\BIBforeignlanguage}[2]{{%
\expandafter\ifx\csname l@#1\endcsname\relax
\typeout{** WARNING: IEEEtran.bst: No hyphenation pattern has been}%
\typeout{** loaded for the language `#1'. Using the pattern for}%
\typeout{** the default language instead.}%
\else
\language=\csname l@#1\endcsname
\fi
#2}}
\providecommand{\BIBdecl}{\relax}
\BIBdecl

\bibitem{Sato77}
H.~Sato, ``{Two-user communication channels},'' \emph{IEEE Transactions on
  Information Theory}, vol.~23, no.~3, pp. 295--304, 1977.

\bibitem{Carleial78}
A.~B. Carleial, ``{Interference channels},'' \emph{IEEE Transactions on
  Information Theory}, vol.~24, no.~1, pp. 60--70, 1978.

\bibitem{carleial1975case}
------, ``{A case where interference does not reduce capacity},'' \emph{IEEE
  Transactions on Information Theory}, vol.~21, p. 569, 1975.

\bibitem{HK81}
T.~S. Han and K.~Kobayashi, ``A new achievable rate region for the interference
  channel,'' \emph{IEEE Transactions on Information Theory}, vol.~27, no.~1,
  pp. 49--60, January 1981.

\bibitem{GSW11bosonic}
S.~Guha, I.~Savov, and M.~M. Wilde, ``The free space optical interference
  channel,'' \emph{accepted for Proceedings of the International Symposium on
  Information Theory, St. Petersburg, Russia}, August 2011, arXiv:1102.2627.

\bibitem{ieee1998holevo}
A.~S. Holevo, ``The capacity of the quantum channel with general signal
  states,'' \emph{IEEE Transactions on Information Theory}, vol.~44, no.~1, pp.
  269--273, 1998.

\bibitem{PhysRevA.56.131}
B.~Schumacher and M.~D. Westmoreland, ``Sending classical information via noisy
  quantum channels,'' \emph{Physical Review A}, vol.~56, no.~1, pp. 131--138,
  July 1997.

\bibitem{winter2001capacity}
A.~Winter, ``{The capacity of the quantum multiple-access channel},''
  \emph{IEEE Transactions on Information Theory}, vol.~47, no.~7, pp.
  3059--3065, 2001.

\bibitem{GRUW01}
A.~J. Grant, B.~Rimoldi, R.~L. Urbanke, and P.~A. Whiting, ``{Rate-splitting
  multiple access for discrete memoryless channels},'' \emph{IEEE Transactions
  on Information Theory}, vol.~47, no.~3, pp. 873--890, 2001.

\bibitem{el2010lecture}
A.~El~Gamal and Y.-H. Kim, ``{Lecture notes on network information theory},''
  January 2010, arXiv:1001.3404.

\bibitem{wilde2011book}
M.~M. Wilde, \emph{From Classical to Quantum Shannon Theory}, 2011,
  arXiv:1106.1445.

\bibitem{S11a}
P.~Sen, ``Sequential decoding for some channels with classical input and
  quantum output,'' 2011.

\bibitem{hayashi2003general}
M.~Hayashi and H.~Nagaoka, ``General formulas for capacity of classical-quantum
  channels,'' \emph{IEEE Transactions on Information Theory}, vol.~49, no.~7,
  pp. 1753--1768, 2003.

\bibitem{GLM10}
V.~Giovannetti, S.~Lloyd, and L.~Maccone, ``Achieving the {Holevo} bound via
  sequential measurements,'' December 2010, arXiv:1012.0386.

\bibitem{FHSSW11}
O.~Fawzi, P.~Hayden, I.~Savov, P.~Sen, and M.~M. Wilde, ``Classical
  communication over a quantum interference channel,'' February 2011,
  arXiv:1102.2624.

\bibitem{sato1981capacity}
H.~Sato, ``{The capacity of the Gaussian interference channel under strong
  interference (corresp.)},'' \emph{IEEE Transactions on Information Theory},
  vol.~27, no.~6, pp. 786--788, 1981.

\bibitem{costa1987capacity}
M.~H.~M. Costa and A.~El~Gamal, ``{The capacity region of the discrete
  memoryless interference channel with strong interference.}'' \emph{IEEE
  Transactions on Information Theory}, vol.~33, no.~5, pp. 710--711, 1987.

\bibitem{sasoglu2008successive}
E.~Sasoglu, ``{Successive cancellation for cyclic interference channels},'' in
  \emph{IEEE Information Theory Workshop, 2008. ITW'08}, 2008, pp. 36--40.

\bibitem{CMGE08}
H.~F. Chong, M.~Motani, H.~K. Garg, and H.~El~Gamal, ``On the {Han-Kobayashi}
  region for the interference channel,'' \emph{IEEE Transactions on Information
  Theory}, vol.~54, no.~7, pp. 3188--3195, 2008.

\bibitem{CT91}
T.~M. Cover and J.~A. Thomas, \emph{Elements of Information Theory}.\hskip 1em
  plus 0.5em minus 0.4em\relax John Wiley \& Sons, 1991.

\bibitem{itit1999winter}
A.~Winter, ``Coding theorem and strong converse for quantum channels,''
  \emph{IEEE Transactions on Information Theory}, vol.~45, no.~7, pp.
  2481--2485, 1999.

\bibitem{ON07}
T.~Ogawa and H.~Nagaoka, ``Making good codes for classical-quantum channel
  coding via quantum hypothesis testing,'' \emph{IEEE Transactions on
  Information Theory}, vol.~53, no.~6, pp. 2261--2266, June 2007.

\end{thebibliography}

\end{document}